\documentclass[journal,12pt,onecolumn,draftclsnofoot]{IEEEtran}
\ifCLASSOPTIONcompsoc
  \usepackage[nocompress]{cite}
\else
  \usepackage{cite}
\fi
\ifCLASSINFOpdf
 \usepackage[pdftex]{graphicx}
\else
  \usepackage[dvips]{graphicx}
\fi
\graphicspath{{figures/}}

\usepackage{amsmath}
\usepackage{mathtools}

\usepackage{amsfonts}

\DeclareMathOperator{\erf}{erf}

\usepackage{upgreek}

\usepackage{array}
\ifCLASSOPTIONcompsoc
  \usepackage[caption=false,font=footnotesize,labelfont=sf,textfont=sf]{subfig}
\else
  \usepackage[caption=false,font=footnotesize]{subfig}
\fi
\usepackage{url}
\newtheorem{proposition}{Proposition}[section]
\newtheorem{corollary}{Corollary}[section]
\newtheorem{definition}{Definition}[section]
\newtheorem{lemma}{Lemma}[section]
\newtheorem{theorem}{Theorem}[section]

\begin{document}
\title{IDEALEM: Statistical Similarity Based\\ Data Reduction}

\author{Dongeun~Lee,
Alex~Sim,
Jaesik~Choi,
and~Kesheng~Wu% <-this % stops a space
\IEEEcompsocitemizethanks{\IEEEcompsocthanksitem D. Lee is with the Department
of Computer Science, Texas A\&M University-Commerce, Commerce,
TX 75428.\protect\\
E-mail: dongeun.lee@tamuc.edu
\IEEEcompsocthanksitem A. Sim and K. Wu are with the Data Science and Technology Department, Lawrence Berkeley National Laboratory, Berkeley, CA 94720.\protect\\
E-mail: asim@lbl.gov, kwu@lbl.gov
\IEEEcompsocthanksitem J. Choi is with the Graduate School of AI, Korea Advanced Institute of Science and Technology, Daejeon 34141, Korea.\protect\\
E-mail: jaesik.choi@kaist.ac.kr}% <-this % stops an unwanted space
\thanks{}}

\IEEEtitleabstractindextext{%
\begin{abstract}
Many applications such as scientific simulation, sensing, and power grid monitoring tend to generate massive amounts of data, which should be compressed first prior to storage and transmission. These data, mostly comprised of floating-point values, are known to be difficult to compress using lossless compression. A few compression methods based on lossy compression have been proposed to compress this seemingly incompressible data. Unfortunately, they are all designed to minimize the Euclidean distance between the original data and the decompressed data, which fundamentally limits compression performance. We recently proposed a new class of lossy compression based on statistical similarity, called IDEALEM, which was also provided as a software package. IDEALEM has demonstrated its performance by reducing data volume much more than state-of-the-art compression methods while preserving unique patterns of data.
IDEALEM can operate in two different modes depending on the stationarity of input data. This paper presents compression performance analyses of these two modes, and investigates the difference between two transform techniques targeted for non-stationary data. This paper also discusses the data reconstruction quality of IDEALEM using spectral analysis and shows that important frequency components in application domain are well preserved. We expand the capability of IDEALEM by adding a new min/max check that facilitates preserving significant patterns lasting only for a brief duration which were previously hard to capture. This min/max check also accelerates the encoding process significantly. Experiments show IDEALEM preserves significant patterns in the original data with faster encoding time.
\end{abstract}}

%\begin{IEEEkeywords}
%Lossy data compression, floating-point data, streaming data, time series data, locally exchangeable measure, online algorithm.
%\end{IEEEkeywords}}

\maketitle

\IEEEdisplaynontitleabstractindextext

% \fi
%\IEEEpeerreviewmaketitle

\IEEEraisesectionheading{\section{Introduction}\label{sec:introduction}}

\IEEEPARstart{C}{omputer} systems and devices generate huge volumes of data, posing a challenge to data management~\cite{reed2015exascale}. Data compression has been considered as common technique for effectively reducing resources required to store and transmit data. Besides, the speed gap between computation and storage widens continuously, which attracts considerable attention to data compression again~\cite{tao2019optimizing,liang2018error,lakshminarasimhan2013isabela}.

Data compression, in a general sense, is based on the idea of leveraging repeated patterns or common features in original data~\cite{cover2012elements}. Depending on the importance of data integrity, a compression method can be designed in either \emph{lossless} or \emph{lossy} way. In lossy compression, we may drop unnecessary or less important information in the original data to gain more reduction in data volume. It is noteworthy that data compression is still considered as an art where a one-size-fits-all approach does not exist: different data types require different solutions.

Among many types of data generated today, floating-point data such as numerical values takes a significant portion. However, floating-point data is known to be especially hard to compress. Although there have been a few publications to losslessly compress floating-point data~\cite{Burtscher:2009:FPC,Lindstrom:2006:FEC}, their compression performance was inherently limited by the characteristics of numerical values that are noisy and random~\cite{lakshminarasimhan2013isabela,Lakshminarasimhan:2011:ISABELA}. Thus it is not surprising that most publications considered lossy compression as their choice on compressing floating-point data~\cite{Iverson:2012:SQE,lakshminarasimhan2013isabela,Lindstrom:2014:ZFP,liang2018error}.

Unfortunately, the quality of all these techniques has been and is still judged by the Euclidean distance between
the original data and the decompressed version of the compressed data. This focus on a single quality measure has imposed a significant limitation on the effectiveness of the compression methods. If we take into account the noisy and random nature of numerical values and the lossy compression, why would we need to restrict ourselves to the Euclidean distance measure? To break this limitation, we propose a new type of compression method based on a
statistical concept known as exchangeability~\cite{lee2016novel,lee2017expanding,lee2017improving,wu2017statistical}. The goal of this compression method is to capture the essential characteristics of data in a statistical sense.

Since we believe the key limitation of the existing compression methods is attempting to reproduce every rise and fall in the original data records, we instead try to preserve key statistical properties of the original data. In particular, we assume a handful of random number generators represented by probability distributions behind actual data, and compress data by capturing only these probability distributions that exist behind. We propose to use the locally exchangeable measure (LEM)~\cite{jaesikLEM} to quantify the similarity of two data blocks. When some blocks are interchangeable, we could keep only a single copy and therefore reduce data volume. By focusing on the statistical properties of data, our approach does not aim to reproduce the original data with small Euclidean distances, but reconstruct data that have the same probability distributions as the original data. This is a significant departure from the common practice in designing compression techniques.

This paper proposes IDEALEM (Implementation of Dynamic Extensible Adaptive Locally Exchangeable Measures) to realize this unique approach~\cite{IDEALEM}. IDEALEM can handle both stationary (including locally stationary) and non-stationary time series, where non-stationary time series data is first handled by transformation methods that can convert non-stationary time series to stationary time series before processed through LEM. In particular, we propose residual and delta transformations to capture long-range trends in data~\cite{lee2017improving,lee2017expanding}. These methods can also work with values in bounded ranges, such as angles between $0^{\circ}$ and $360^{\circ}$. We demonstrate the effectiveness of IDEALEM through an extensive evaluation.
% distribution within each block is stationary and the delta transformation assumes the incoming values are generated with a Gaussian random walk
% a set of power grid monitoring data, a data set from neuroscience

The main contributions of this paper, compared with our previous work~\cite{lee2016novel,lee2017expanding,lee2017improving,wu2017statistical}, are as follows:
\begin{itemize}
\item We provide a comprehensive analysis on the compression performance of IDEALEM in two different modes targeted for stationary and non-stationary data.
\item We closely analyze the residual and delta transformations for non-stationary data in terms of exchangeability and show they have different characteristics.
\item Spectral analysis shows the quality of reconstructed data is not compromised for application domain considering the frequency components of original and reconstructed data.
\item To complement the statistical similarity, we employ a min/max check that can better preserve significant patterns lasting only for a brief duration. This check also reduces execution time for the encoding operation.
\end{itemize}
% \item We propose a simple way to match the minimum and maximum values of
%   the data blocks before performing the KS~test.  This has three
%   important consequence.  It is much easier to implement than
%   statistical similarity tests that are sensitive to extreme values.  It
%   also reduces the number of KS~tests performed, which reduces the
%   execution time for the compression operation.  More importantly, this
%   simple test allows us to identify blocks with extreme values as unique
%   blocks and therefore preserving them as important features for further
%   analysis.

The rest of this paper is organized as follows. Section~\ref{sec:related} reviews related work. In Section~\ref{sec:idealem}, we discuss a statistical similarity measure that can be used in conjunction with the LEM concept and the key design considerations of our algorithm.  We present the details of IDEALEM implementation in Section~\ref{sec:operation} and its encoding and decoding processes in Section~\ref{sec:encoding}. Section~\ref{sec:theoretical} provides theoretical analysis on the achievable compression ratios of IDEALEM, residual/delta transformation in terms of exchangeability, and the effect of duplicating data blocks in decoding. Section~\ref{sec:experimental} reports various experimental results of IDEALEM and discusses its performance, followed by concluding remarks in Section~\ref{sec:conclusion}.

%\IEEEpubidadjcol

\section{Related Work} \label{sec:related}
Data compression is a way of representing information in a compact form. This is accomplished by identifying and using structures that exist in the data~\cite{sayood2012introduction}. A data compression model is categorized into two broad classes: \textbf{lossless coding} where a reconstruction of compressed data is identical to the original raw data; and \textbf{lossy coding} where a reconstruction is different from the original raw data, while providing much higher compression. A choice of compression model should take account of data type. If keeping the integrity of data is of utmost importance, then the lossless coding should be selected for the compression. On the other hand, if one can tolerate a certain amount of distortion and approximate results are sufficient most of the time~\cite{srisooksai2012practical,lakshminarasimhan2013isabela,Lakshminarasimhan:2011:ISABELA}, the quality of data can be adjusted in favor of better compression with the lossy coding~\cite{lee2015sensors,lee2015scalable,lee2014big,lee2015learning}.

\subsection{Related Lossless Coding Schemes}
Lossless coding methods similar to IDEALEM lie in the area of biological sequence compression, especially deoxyribonucleic acid (DNA) compression~\cite{cao2007simple}. Some DNA sequences are highly repetitive, but they are not exactly identical to the original sequence as nucleotides can be changed, inserted, or deleted. This is the reason why most conventional dictionary-based algorithms~\cite{ziv1977universal,ziv1978compression,welch1984technique,sayood2012introduction} fail to compress DNA data, as they all try to look up the same recurring pattern stored in the dictionary. To handle this, substitution approaches exploiting approximate repeats have been proposed~\cite{chen2000compression,chen2002dnacompress}. IDEALEM also reconstructs data from learned patterns during the encoding process that do not need to be identical. However, IDEALEM adopts the lossy coding in contrast to DNA compression, as it targets for quality-adjustable data.

\subsection{Related Lossy Coding Schemes and Their Limitations} \label{sec:lossy}
On the other hand, conventional lossy coding schemes in general quantize or threshold data to adjust quality and reduce data size~\cite{sayood2012introduction}. Their goal is to compress data without compromising distinctive attributes of data. However, the tenets of these conventional schemes thus far have restricted their attention to the recovery of signal where distortion (distance) is measured using $\ell_{2}$ norms (Euclidean distance) such as mean squared error (MSE) and signal-to-noise ratio (SNR)~\cite{richardson2011h,candes2008introduction,sayood2012introduction,lee2014big}. Specifically, using $\ell_{2}$ norms as the distance measure requires the sequence of encoded and decoded data to be preserved. IDEALEM relaxes this constraint and instead treats a sequence of data as if it originates from random numbers, which eventually leads to superior compression ratio~\cite{lee2016novel}.

Employing the concept of random numbers introduces a new way of signal recovery: data is reconstructed from a learned probability distribution during the encoding process, not from the encoded (quality-adjusted) data itself. Thus, the encoded output is not a direct representation of the original data; instead, the encoder informs the decoder how to regenerate them. From the perspective of data compression, this approach can be regarded as one of analysis/synthesis schemes~\cite{sayood2012introduction}. In particular, IDEALEM resembles the concept of fractal compression~\cite{jacquin1992image,wohlberg1999review} in the sense that both rely on the self-similarity of data, which suggests that parts of data often resemble other parts of the same data.

Nevertheless, the goal of fractal compression is still to reconstruct data as close as possible to original data in terms of Euclidean distance. To this end, similar data part found should be transformed in order to closely match a target data part, which incurs lots of computational overhead. IDEALEM parts with the conventional Euclidean distance measure; it rather focuses on the exchangeability of similar data sequences. In particular, this flexibility on the order of data sequence not only yields high compression ratio, but makes the encoding process of IDEALEM faster than the fractal compression.

Another work related to IDEALEM is \textbf{clustering}~\cite{aggarwal2003framework,guha2003clustering,tabata2010data,jain2010data}. Clustering is to group data samples such that similar samples are in the same group and different samples are in different groups. One of the most popular clustering algorithm is $k$-means clustering whose similarity is typically defined using the Euclidean distance measure~\cite{jain2010data}. $k$-means is also closely connected with data compression, where it is called vector quantization~\cite{sayood2012introduction,linde1980algorithm}. Since IDEALEM keeps only a single copy of data block among interchangeable data blocks to reduce data volume, it can be essentially considered as a new type of vector quantization with statistical similarity. Similarly, IDEALEM can be also used as a clustering method.

\subsection{Floating-Point Compression} \label{sec:floating}
From the viewpoint of applications, IDEALEM compresses the one-dimensional array of floating-point values. The current state-of-the-art compression algorithms for floating-point data are \textbf{ZFP}~\cite{Lindstrom:2014:ZFP,LindstromZFP}, \textbf{ISABELA}~\cite{Lakshminarasimhan:2011:ISABELA,ISABELASW}, and \textbf{SZ}~\cite{liang2018error,SZSW}. ZFP adopts lossy coding and its design is also based on the Euclidean distance measure for the recovery of signal. ZFP can compress up to three-dimensional floating-point arrays. It is reported that ZFP can achieve compression ratios on the order of 100 for three-dimensional arrays with higher quality than other compression algorithms, because it can exploit strong correlation among three dimensions. However, we can see the performance of ZFP suffers with one-dimensional array~\cite{lee2016novel}.

ISABELA (In-situ Sort-And-B-spline Error-bounded Lossy Abatement) applies data sorting to change irregular data pattern to a smooth and monotonous curve, which is further fitted by the B-spline. Adopting lossy coding, ISABELA can achieve data compression by storing only B-spline constants: knot vector and basis coefficients, along with difference between B-spline estimated and actual values.\footnote{ISABELA is originally targeted for spatio-temporal data (two-dimensional arrays), where the spatial dimension is handled by sorting and B-spline; the temporal dimension by delta encoding of index values~\cite{Lakshminarasimhan:2011:ISABELA}. However, its available implementation~\cite{ISABELASW} only supports the spatial dimension processing, which we regard as the one-dimensional array processing in this paper.} However, ISABELA is also grounded on the Euclidean distance measure. Thus it has to record the index of sorted data values for associating decoded data with their original ordering, which incurs an additional overhead. The performance of ISABELA suffers from both low compression ratio and slow encoding time.

SZ (squeeze) is a recently proposed lossy coding scheme that shows superior performance. Similar to ISABELA, SZ employs curve fitting; but it provides three curve-fitting models (preceding neighbor fitting, linear-curve fitting, and quadratic-curve fitting) and chooses the one that best predicts each target value. For unpredictable data values by these three models, SZ analyzes their binary representation to compress them anyway. While SZ shows superior performance and in most cases performs better than ZFP, it is still based on the Euclidean distance measure, which fundamentally limits compression performance.

\section{Implementation of Dynamic Extensible Adaptive Locally Exchangeable Measures} \label{sec:idealem}
Various application scenarios that generate floating-point values can be explained by random number generations: devices such as sensors might be measuring background noise during their operation time, and network monitoring devices would be observing random traffic. We demonstrate the performance of IDEALEM on these data by employing LEM, which relaxes the order of data sequence and therefore is very effective in terms of compression performance.

Some of these data, however, are not immediately compressible by simply treating them as random numbers. Of particular interest is non-stationary data such as the phase angle of electricity data that has a constantly increasing trend. For this type of data, we introduce two transformation methods that can locally get rid of non stationarity: residual and delta transformations, where residual values with reference to the base value or differences between sequential values are treated as random numbers, respectively. These two transformation methods are shown to have good performance for non-stationary data.

\begin{figure}[!t]
\centering
\includegraphics[width=0.5\columnwidth]{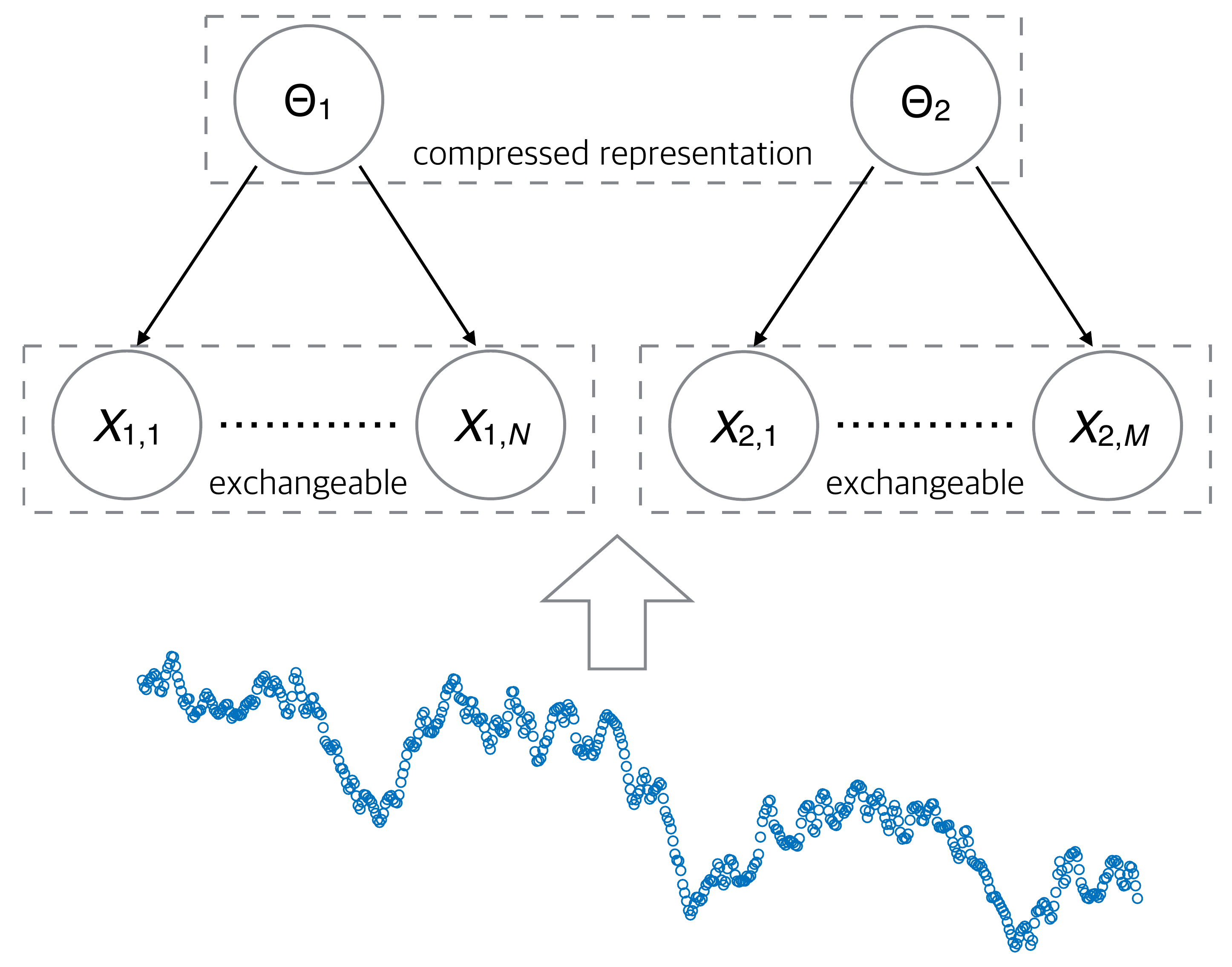}
\caption{IDEALEM breaks an input data into fixed-size blocks and treats each of them as an instantiation of a random variable. IDEALEM compresses the input data by learning common probability distributions behind two groups of random variables, which are represented by latent random variables $\Theta_1$ and $\Theta_2$. These distributions are non-parametric, allowing any shapes of distributions. In this example, $\Theta_1$ governs the identical distribution of $N$ random variables $X_{1,i}$; $\Theta_2$ the identical distribution of $M$ random variables $X_{2,j}$. As a result, $N+M$ data blocks are reduced to only 2 data blocks.}
\label{fig_bayes}
\end{figure}

The main idea of IDEALEM is to store only data sequences that are distinct from previous data sequences in terms of statistical similarity. To this end, IDEALEM breaks an incoming data stream into blocks of a fixed size and represents statistically similar blocks with a data block that appears earlier in the data stream. If we assume that each data block is an instantiation of a random variable, we can consider an exchangeability of these random variables, where the exchangeability can be assumed if these random variables share an identical distribution as their data source, as described in Fig.~\ref{fig_bayes}.\footnote{Note that the term \emph{exchangeability} here is used in somewhat wide sense. Rigorously speaking, random variables having the identical distribution are not necessarily exchangeable, although the converse is true~\cite{jaesikLEM}.}

Since the exchangeable random variables shown in Fig.~\ref{fig_bayes} can be represented with any of them, IDEALEM stores the first occurrence of these exchangeable random variables in streaming time series, and represents others using the stored data block. Therefore, if there are more of similar random variables, we can achieve a higher compression ratio. Here, the compression ratio is defined by the ratio of the original data size to the compressed size.

\subsection{Similarity Measure Based on Kolmogorov-Smirnov Test} \label{sec:similarity}
Our statistical similarity measure is especially based on Kolmogorov-Smirnov test (KS~test)~\cite{yu2005ovarian, seabra2008modeling, quinsac2010compressed}, which is by far the most popular test for checking the similarity of time series. Moreover, KS~test is simple to implement and faster than other non-parametric statistical hypothesis testing methods such as Anderson-Darling test~\cite{engmann2011comparing}. Essentially, the purpose of KS~test is to test whether two underlying one-dimensional probability distributions of random variables differ or not. Being a non-parametric test, the KS~test can compare two random variables from any arbitrary distributions without the restriction of parametric distribution assumption.

\begin{figure}
\centering
\includegraphics[width=0.5\columnwidth]{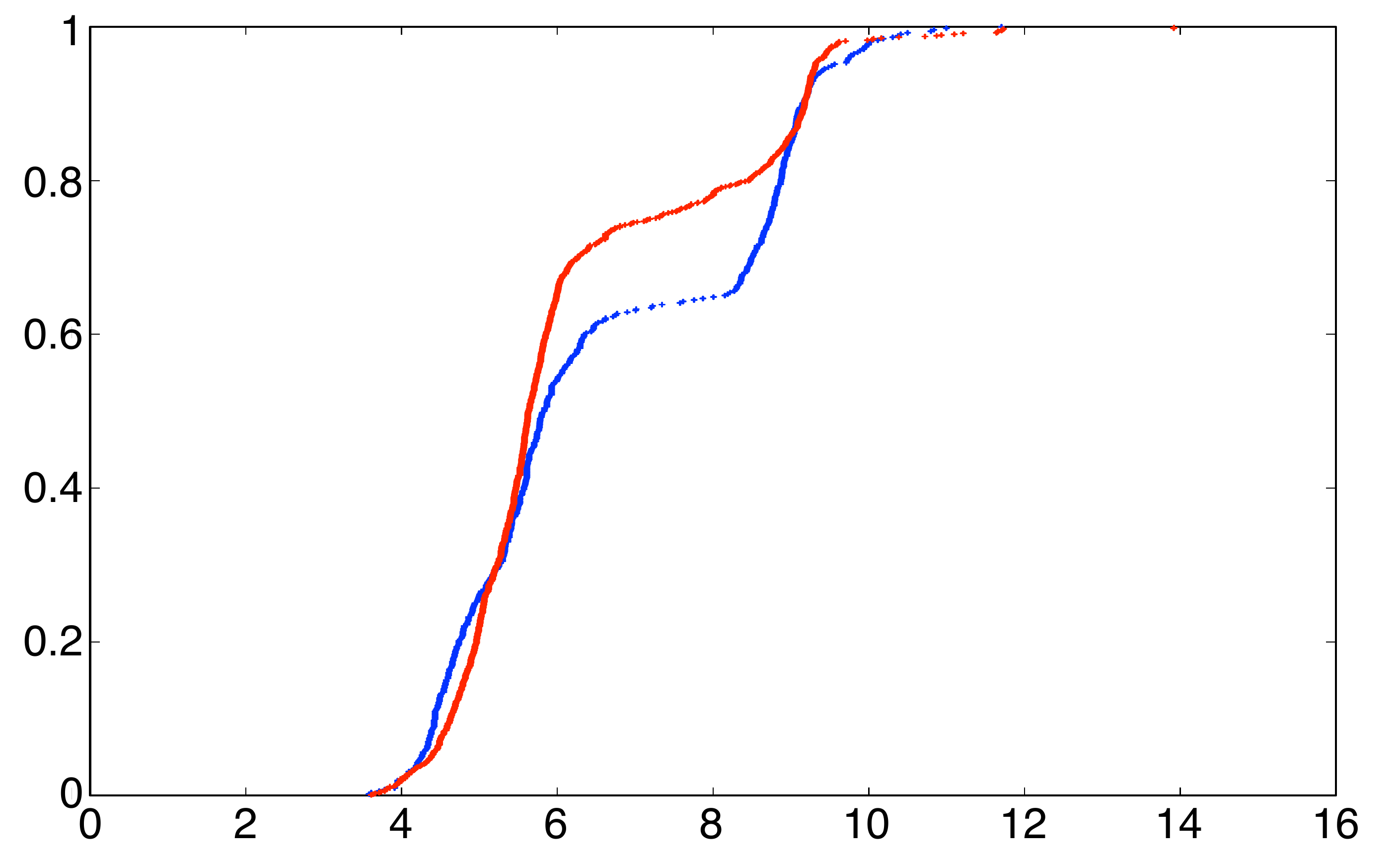}
\caption{An example of two empirical distributions in blue and red
  colors.  The distributional distance is visible in the
  middle.  The maximum distance, defined by (\ref{eq_one}), ranges between zero and one.}
\label{fig_KS}
\end{figure}

Basically, KS~test computes the maximum distributional distance between two random variables, i.e., the two-sample KS~test, and this distance is \emph{standardized} according to the numbers of samples for two random variables. Here, the maximum distributional distance $D_{n_i,n_j}$ between two random variables $X_i$ and $X_j$ is defined by
\begin{equation}
\label{eq_one}
D_{n_i,n_j}\coloneqq\begin{subarray}{c}
\mathrm{\displaystyle sup}\\
x
\end{subarray}\, |F_{X_i,n_i}(x)-F_{X_j,n_j}(x)|,
\end{equation}
where $F_{X_i,n_i}(\cdot)$ and $F_{X_j,n_j}(\cdot)$ are empirical
(cumulative) distribution functions of $X_i$ and $X_j$; $n_i$ and $n_j$
are the numbers of samples for $X_i$ and $X_j$ respectively; $\sup$ is
the supremum. Fig.~\ref{fig_KS} shows an example of two empirical
distributions where we can clearly see a distributional distance between
them. The distance (\ref{eq_one}) is called the \textbf{test
  statistic}, and is subsequently standardized with respect to
$n_i$ and $n_j$ as follows:
\begin{equation}
\label{eq_two}
D_{n_i,n_j}\sqrt{\frac{n_{i}n_{j}}{n_{i}+n_{j}}}.
\end{equation}

The larger this standardized distance is, the smaller the output of the KS~test is, which is called the \textbf{p-value}~\cite{wasserstein2016asa}. A high p-value indicates that two random variables are likely to be from the same distribution; while a small p-value indicates that they are less likely to be from the same distribution. In IDEALEM, we use the p-value as the indicator of identifying the exchangeability.

It should be noted that given two same maximum distributional distances between random variables, the standardized distances (\ref{eq_two}) of them could be different depending on the numbers of samples for the random variables, due to the scaling factor $\sqrt{n_{i}n_{j}/(n_{i}+n_{j})}$. For instance, this factor is simply $\sqrt{n/2}$
when $n_i=n_j=n$. As a result, a larger number of samples for each random variable (i.e., for each data block) tends to produce a smaller p-value given the same maximum distributional distance. This attribute of the KS~test, which we call \emph{sensitivity with number of samples}, affects compression performance in the sense that although a large number of samples for each data block potentially increases compression ratio, which is proven in Section~\ref{sec:fundamental}, it also increases the difficulty of passing the KS~test due to the sensitivity with the number of samples, which causes an adversarial effect for the compression performance.

\begin{figure}
\centering
\includegraphics[width=0.6\columnwidth]{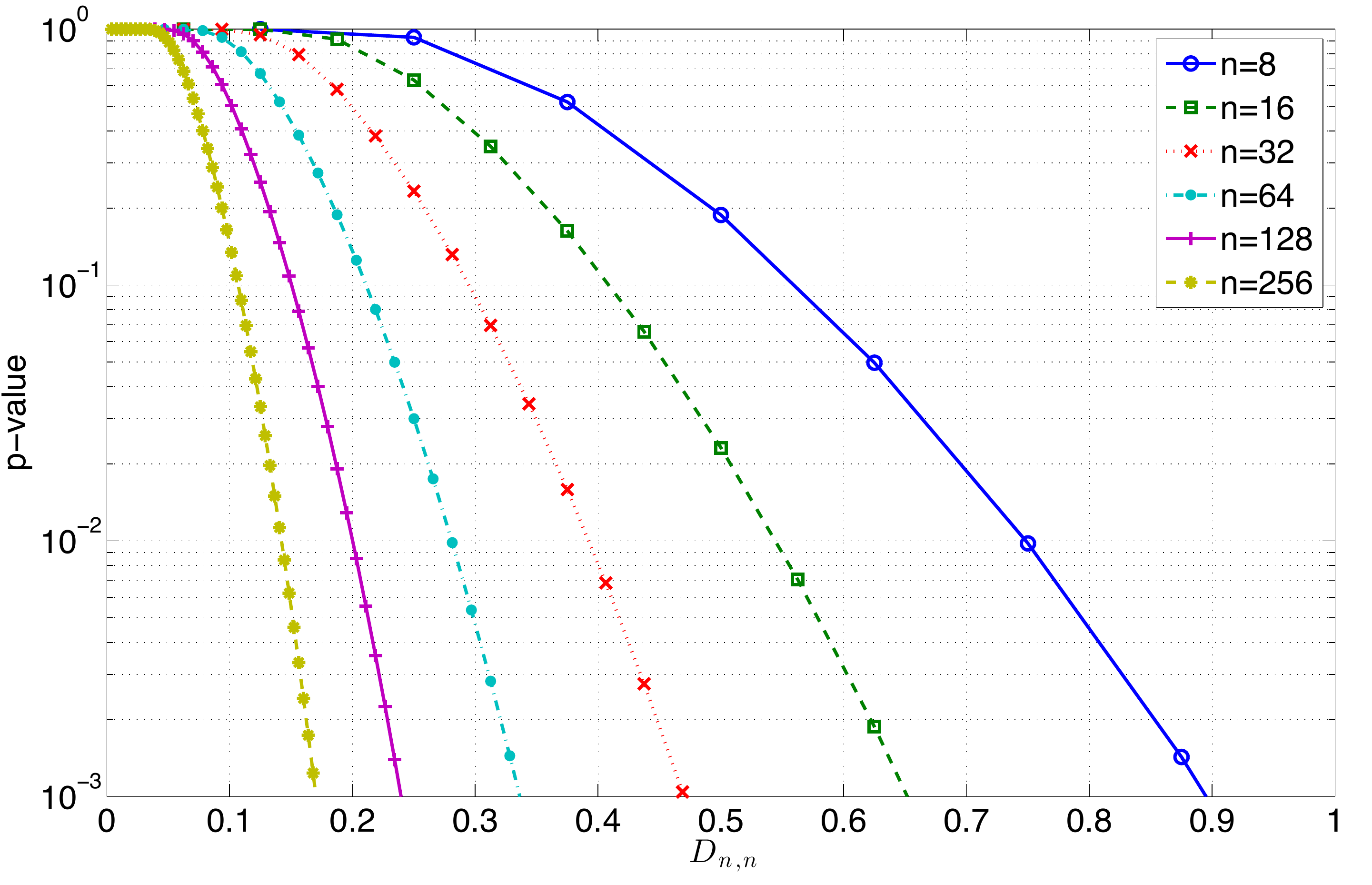}
\caption{Effects of numbers of samples on the p-value and corresponding
  test statistics (maximum distributional distances) $D_{n,n}$, where the y-axis is drawn in log scale.
  Six different cases are synthetically
  generated, where each has $n$ samples for two random variables.
  As $n$ grows, it becomes more difficult to exchange random variables due to
  lower p-values for a given distance.}
\label{fig_three}
\end{figure}

To visualize this sensitivity, Fig.~\ref{fig_three} shows the plot of the p-value versus the test
statistic (\ref{eq_one}) with various $n$'s ($n_i=n_j=n$).  Given the maximum distributional distance
$D_{n,n}$, a larger $n$ leads to a smaller p-value.  Thus even a small
distance with a large $n$ could lead to a small p-value.  In other
words, the same p-value may correspond to different test statistics
depending on $n$.  Since this sensitivity obviously affects compression performance, we can find an optimum number of samples for a given data set, for which we recently proposed an online algorithm~\cite{gibson2018dynamic}.

\subsection{Min/Max Check with Relative Tolerance} \label{sec:minmax}
Previously, IDEALEM had difficulties in preserving significant patterns that lasted for a very brief duration~\cite{lee2016novel}. In particular, a sudden increase or decrease in value was hard to capture due to the insensitivity of KS~test to differences in the tails of distributions. Although this problem could be somehow mitigated with a combination of parameters of IDEALEM, the interplay among these parameters was so complex that there was no general principle.

In order to resolve this issue of preserving significant patterns, this paper proposes a simple, yet effective way that can complement KS~test. Since the KS~test decides that two data sequences (random variables) share the same underlying probability distribution (i.e., two random variables are exchangeable) as far as the majority of samples from two sequences has similar values, differences in values that are larger or smaller than the majority of values are by nature neglected. Unfortunately, significant patterns that only last for a brief duration coincide with these differences in both tail parts of the distribution.

We check if minimum and maximum values of a data sequence fall on tolerable ranges calculated with reference to low and high ends of other data blocks stored in buffers, where the tolerable range can be increased or decreased according to a relative tolerance parameter. Placed before the KS~test routine, this min/max check effectively filters out data blocks that contain significant patterns for brief durations. In addition, this check accelerates the encoding process of IDEALEM by helping the encoding process to bypass the KS~test routine whenever it is unnecessary: it detects and filters out unexchangeable data sequences early.

\begin{figure}
\centering
\includegraphics[width=0.5\columnwidth]{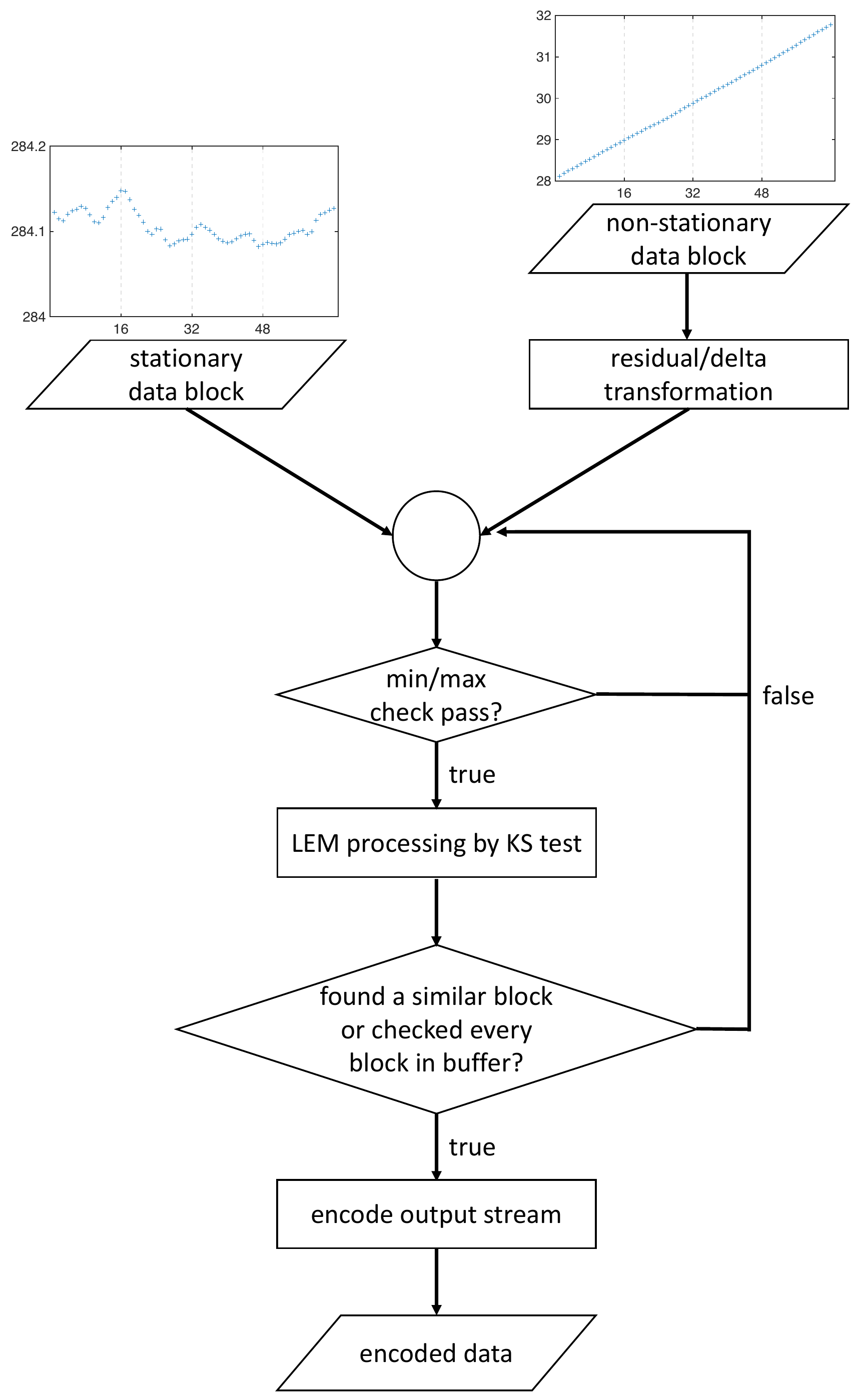}
\caption{Flowchart for the encoding process of IDEALEM. Learned probability distributions are stored in the buffer, which is searched for a similar block until it is exhausted.}
\label{fig_overview}
\end{figure}

Fig.~\ref{fig_overview} overviews the encoding process of IDEALEM. Non-stationary data should first go through the residual or delta transformation before processed further. The min/max check checks if each data block lies on the tolerable ranges. Only after that is the data block processed using LEM by KS test. IDEALEM maintains a buffer during the encoding process where it stores learned probability distributions $\Theta$ in the form of data blocks. If a statistically similar data block is found in the buffer during the search, IDEALEM only outputs an index to the existing similar block. Otherwise, when the incoming data block is statistically different from all others, it is stored in the buffer and also outputted precisely as is.

\section{Operation of IDEALEM} \label{sec:operation}
IDEALEM provides a few parameters that serve as the tuning knobs of its operation. \textbf{Block size} $B$ determines the number of samples in an individual data block. An incoming time-series data is broken down into blocks with each of them having $B$ elements. The block size has an interesting effect on compression performance due to the combination of positive and negative effects on the compression performance, as explained in Section~\ref{sec:similarity}.

\textbf{Number of dictionary blocks} $D$ controls how many source distributions are stored in buffer memory for comparison, where each dictionary block holds one probability distribution. The number of dictionary blocks plays an important role in compression performance: more dictionary blocks in general promise higher compression ratios because there is a higher chance of finding a similar distribution stored in the buffer when we process a new data block. In other words, the number of dictionary blocks affects the locality of LEM processing.

However, increasing $D$ may have drawbacks: (i) we cannot simply store too many dictionary blocks at the same time in memory, especially for resource-limited devices such as sensors; (ii) a new data block is compared against each distribution stored in memory to compute the p-value, which could potentially increase the execution time especially when it is difficult to find a similar data block, because the more dictionary entries we keep, the more KS~tests are performed. However, since IDEALEM currently limits $D$ to the maximum of 255, memory consumption is not an issue unless $B$ is extremely large. Besides, we introduce the min/max check in this paper, and this check greatly reduces the encoding time of IDEALEM by helping the encoding process to bypass the KS~test.

\textbf{Threshold} $\alpha$ is the threshold for similarity when comparing a new data block to source distributions stored in buffers using the KS~test.\footnote{This value is called the \emph{significance level} in statistics literature. With a chosen $\alpha$, we can only say the \emph{type I error} rate is at most $\alpha$, which is the incorrect rejection of a true null hypothesis: two random variables are from the same distribution. IDEALEM utilizes $\alpha$ in a broader sense as a way of identifying similar random variables.} A lower $\alpha$ results in a higher compression ratio, allowing more data blocks to be declared exchangeable and thus to be represented by one of the source distributions stored in the buffer. However, lowering the bar for similarity may impair reconstruction quality, as it would also include not-so-similar sequences under the same source distribution.

\textbf{Relative tolerance} $r$ determines the tolerable ranges of both ends of a source distribution stored in the buffer. The minimum and maximum values of a new data sequence should fall on both tolerable ranges in order to proceed to the next LEM processing by KS test, which we call the min/max check. In particular, if we assume that each new data block is represented by a random variable $X$, and each source distribution by a random variable $\Theta$, the following conditions should be met:
\setlength{\arraycolsep}{0.0em}
\begin{eqnarray}
\label{eq_three}
\min{\Theta}-wr\leq&{}\min{X}{}&\leq\min{\Theta}+wr,\nonumber\\
\max{\Theta}-wr\leq&{}\max{X}{}&\leq\max{\Theta}+wr,
\end{eqnarray}
\setlength{\arraycolsep}{5pt}%
where $w=(\max{\Theta}-\min{\Theta})$ is the width of a source distribution with which the relative tolerance is calculated. In (\ref{eq_three}), if either of conditions is not fulfilled, $X$ is filtered out before the KS~test because it may contain outliers which are difficult to capture using the KS~test (i.e., a significant pattern that only lasts for a brief duration), or it would not pass the KS~test with respect to $\Theta$ anyway due to its range of values being far from the range of $\Theta$.

It should be noted that in (\ref{eq_three}), the upper bound of the first condition ($\min{\Theta}+wr$) and the lower bound of the second condition ($\max{\Theta}-wr$) touch, when $r=0.5$. Therefore, if $r\geq 0.5$, theoretically the range of $X$ could be as small as zero, which means all the values of $X$ are concentrated on a single value.\footnote{Equivalently, we only have a single point mass in this case.} However, in this case, it would be very difficult for $X$ to pass the KS~test.

\subsection{Residual and Delta Transformations} \label{sec:transformations}
The LEM processing of IDEALEM declares two data blocks to be similar when their empirical distributions are close to each other. Assuming data blocks are generated by random processes, compressible time series must be generated from stationary processes. For those time series from non-stationary processes, we seek to transform the data
blocks so that we can use the LEM processing by the KS~test.

Residual transformation and delta transformation are provision for handling non-stationary time series, which manipulate each data block in the following way: the first sample value is designated as a base value, and residual and delta values of other samples are computed. Residual values are computed by subtracting the base value from other sample values; whereas delta values by subtracting one value from the next subsequent value.

Formally speaking, let $x_i$ denote the incoming values ($i=0,1,2,\ldots$). Then the block $j$ ($j=0,1,2,\ldots$), denoted by $b_j$, would include values ($x_{jB}, x_{jB+1}, \ldots, x_{jB+B-1}$). The residual transformation records the base value $x_{jB}$ of $b_j$ and transforms the remaining values of the block as follows:
\begin{equation}
\label{eq_four}
x^r_{jB+k}\coloneqq x_{jB+k}-x_{jB},
\end{equation}
where $k=1,2,\ldots,B-1$. In this case, the LEM processing is performed on
\begin{equation}
\label{eq_five}
b^r_j\coloneqq(x^r_{jB+1},x^r_{jB+2},\ldots,x^r_{jB+B-1}).
\end{equation}
In most non-stationary processes, we expect a probability distribution within $B-1$ values to be relatively stable; therefore, it is possible that different $b^r_j$'s could be similar to each other.

On the other hand, the delta transformation computes successive differences among $x_i$ to make resulting output follow a stationary process. Again, using $x_{jB}$ as the base value of a block $b_j$, the delta transformation computes a new block consisting of
\begin{equation}
\label{eq_six}
x^d_{jB+k}\coloneqq x_{jB+k}-x_{jB+k-1},
\end{equation}
where $k=1,2,\ldots,B-1$. In this case, the LEM processing is performed on
\begin{equation}
\label{eq_seven}
b^d_j\coloneqq(x^d_{jB+1},x^d_{jB+2},\ldots,x^d_{jB+B-1}).
\end{equation}
As in the case of the residual transformation, it is possible that different $b^d_j$'s could be similar to each other.

\begin{figure*}[!t]
\centering
\subfloat[Original]{\includegraphics[width=0.33\linewidth]{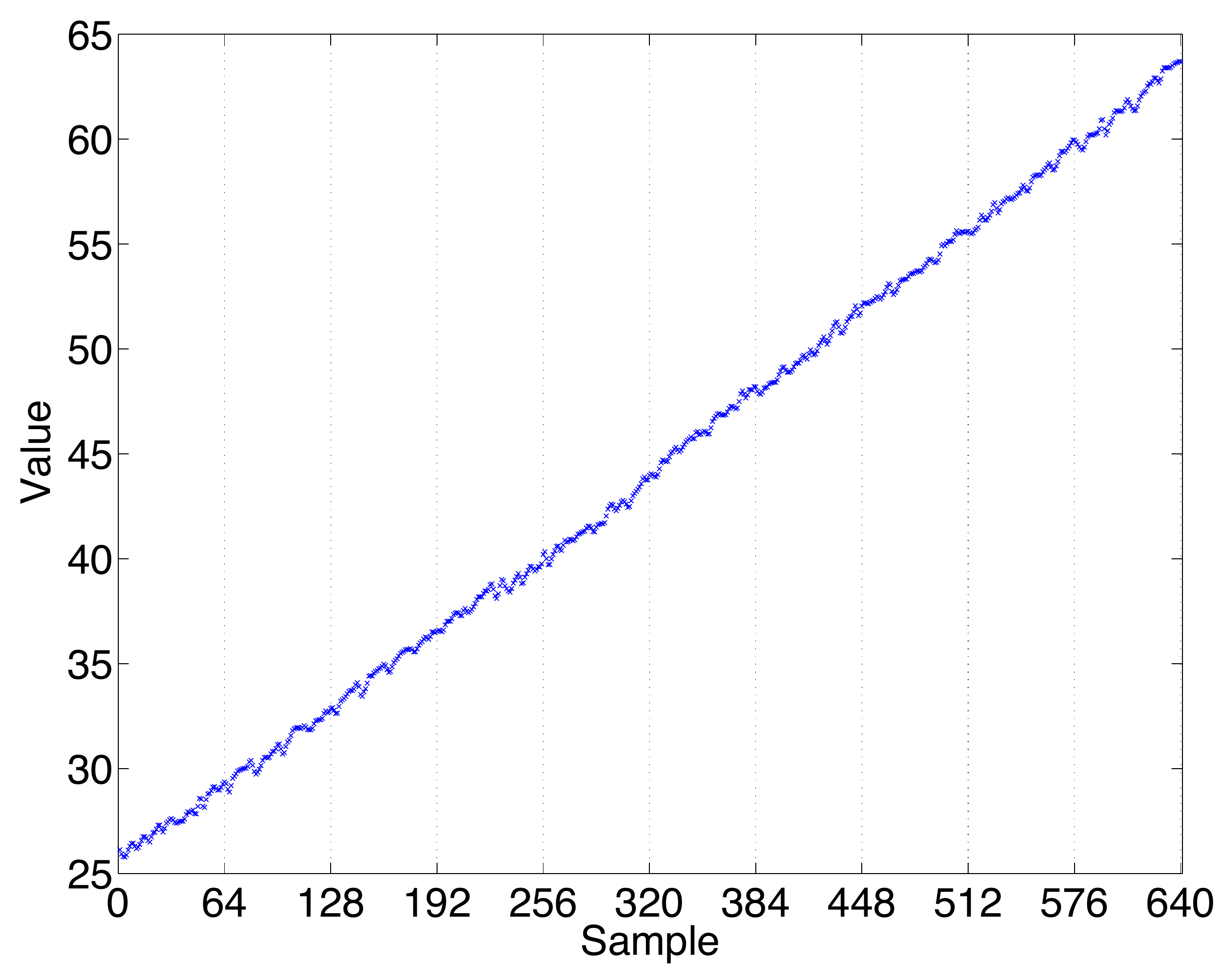}%
\label{fig_five_a}}
\hfil
\subfloat[Residual]{\includegraphics[width=0.33\linewidth]{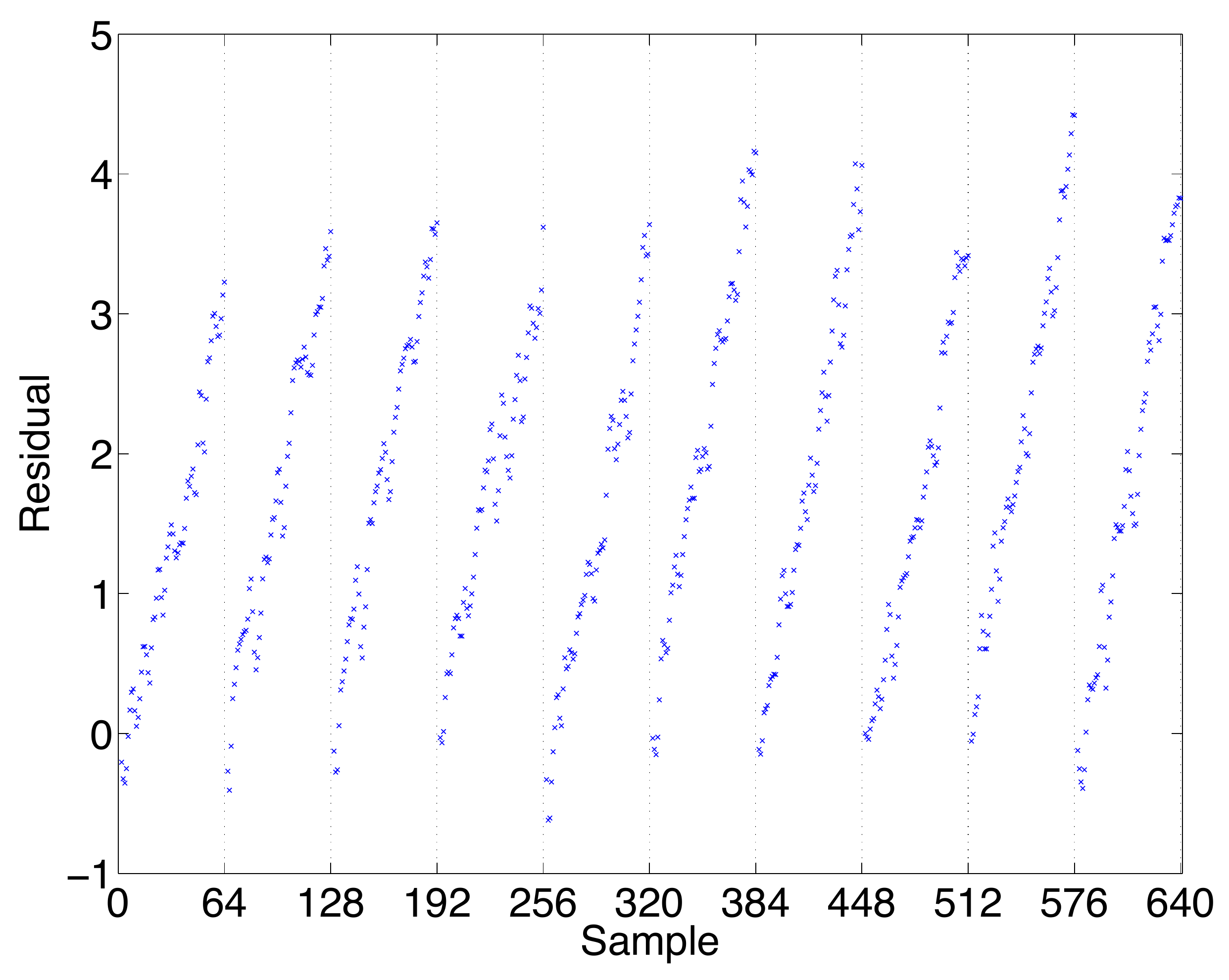}%
\label{fig_five_b}}
\hfil
\subfloat[Delta]{\includegraphics[width=0.33\linewidth]{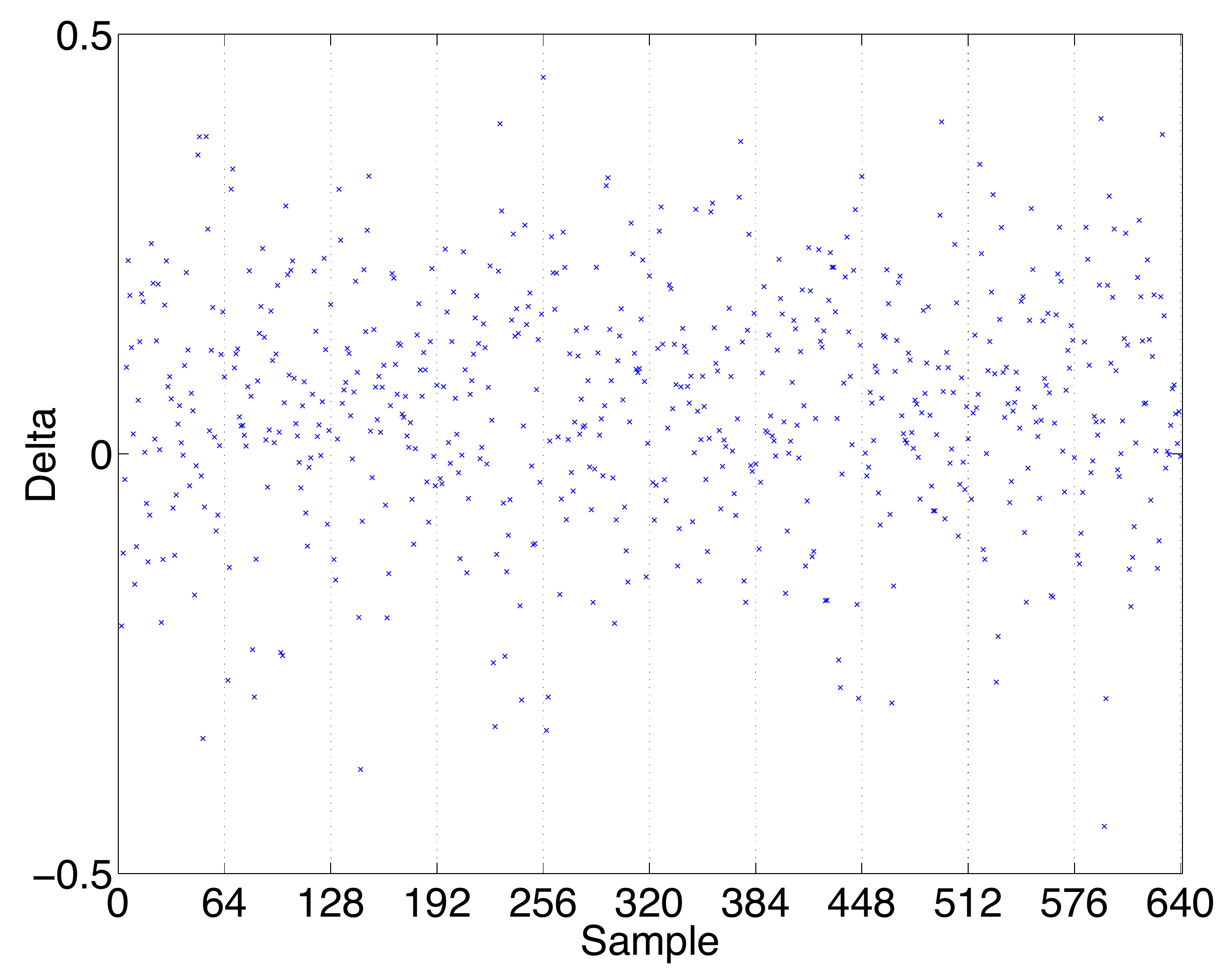}%
\label{fig_five_c}}
\caption{Scatter plot of 640 samples of non-stationary data from power grid electricity data set with the block size $B=64$. Residual values of data blocks are shown in (b), and delta values in (c), both without base values (63 values for each data block). Each data block becomes exchangeable with other blocks in (b) and (c).}
\label{fig_five}
\end{figure*}

Fig.~\ref{fig_five} shows the residual and delta transformations of 640 sample values. The original values in Fig.~\ref{fig_five_a} are phase angle values of electricity data that are difficult to directly compress using the LEM processing. However, after the transformations, which are shown in Fig.~\ref{fig_five_b} and Fig.~\ref{fig_five_c}, new data blocks appear more likely to be exchangeable with each other, and therefore they are more compressible.

\begin{figure}
\centering
\includegraphics[width=0.5\columnwidth]{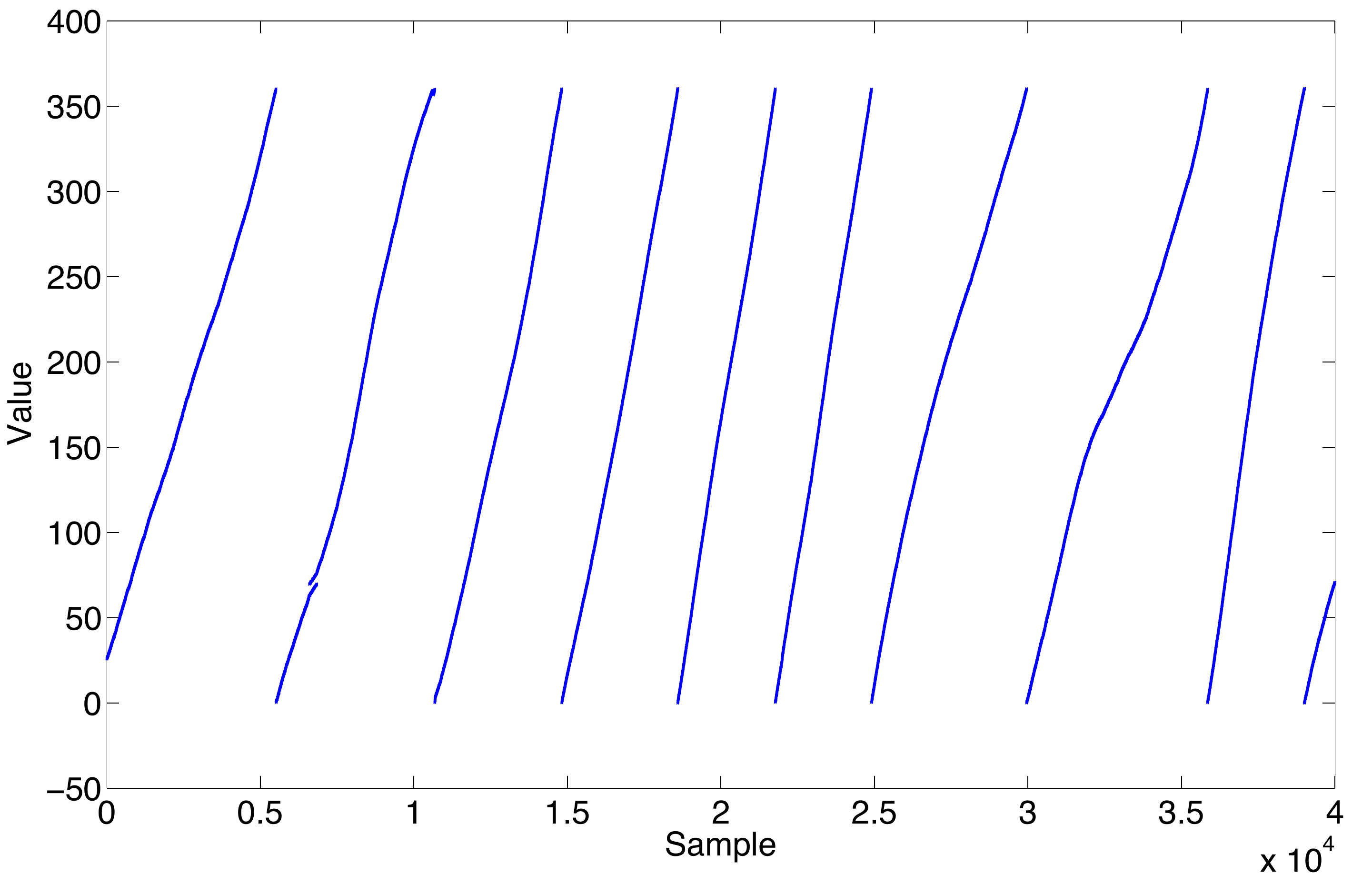}
\caption{Scatter plot of 40,000 samples of phase angle measurements from power grid electricity data set. Values are increasing from $0^{\circ}$ to $360^{\circ}$, where they wrap to $0^{\circ}$ again.}
\label{fig_six}
\end{figure}

Values such as the phase angles of alternating current shown in Fig.~\ref{fig_five_a} may have bounded ranges since they constantly increase: the phase angle measurements have the range of $0^{\circ}$ to $360^{\circ}$, where the value wraps to $0^{\circ}$ once it reaches $360^{\circ}$. Fig.~\ref{fig_six} shows the phase angle measurements for an extended period, where we can observe the periodic nature. In this case, the encoding and decoding processes of a compression method should respect these ranges.

In order to handle the values in bounded ranges, IDEALEM has a provision for controlling the range of encoded and decoded data in the residual and delta transformations. \textbf{Range minimum} $r_{\mathrm{min}}$ and \textbf{range maximum} $r_{\mathrm{max}}$ are two parameters that define the minimum and maximum values of a variable. Once these parameters are set, the encoding process assures that residual/delta values are within the range of $-(r_{\mathrm{max}}-r_{\mathrm{min}})/2$ to $(r_{\mathrm{max}}-r_{\mathrm{min}})/2$, and adjusts values outside of this range so that they fall within the range. For example, a delta value of $359^{\circ}$ and $1^{\circ}$ equals $-358$, which should be adjusted to $2$. On the other hand, the decoding process has to assure that reconstructed data is within the range of $r_{\mathrm{min}}$ to $r_{\mathrm{max}}$, and wrap all values outside of the range to be within the range.

\section{Encoding and Decoding} \label{sec:encoding}
The encoding and decoding processes of IDEALEM can be explained with two separate compression modes. The standard mode is intended for the compression of locally stationary random fluctuations of data, whereas the residual/delta mode is intended especially for the compression of non-stationary data. These two different examples are shown in Fig.~\ref{fig_overview}. The basic idea behind IDEALEM is to store only a source distribution that is distinct from previous source distributions stored in the buffer, according to the statistical similarity measure based on KS~test.

\subsection{Standard Mode} \label{sec:encoding_standard}
\begin{figure}
\centering
\includegraphics[width=0.4\columnwidth]{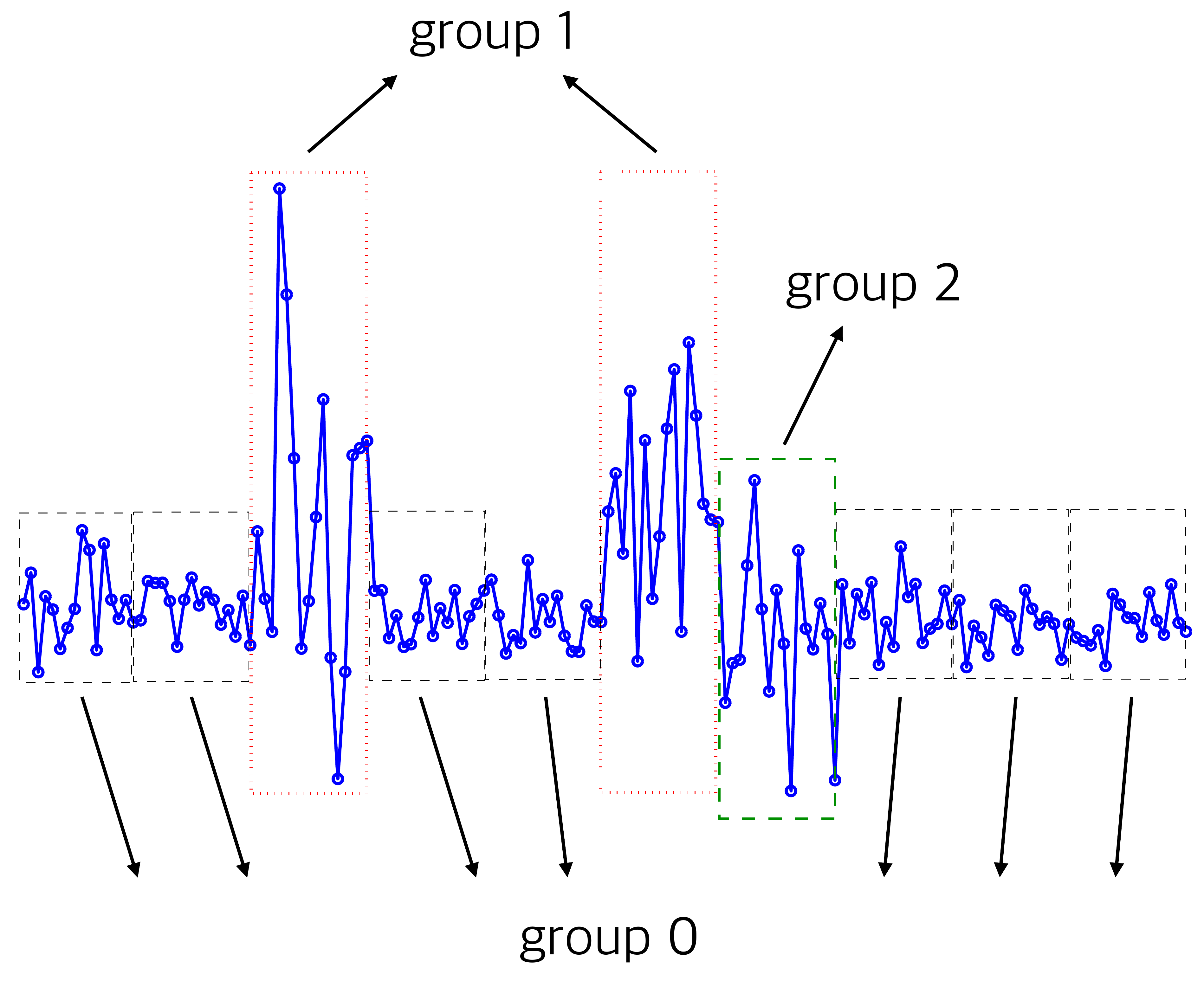}
\caption{An example time-series data of total 160 samples and data blocks with $B=16$. Group 0 and group 1 data blocks are exchangeable with each other within the group. However, the data block in group 2 is unexchangeable and stored separately since it does not have any similar blocks that have appeared earlier in this time series. It might be exchangeable with other data blocks in the future.}
\label{fig_standard}
\end{figure}

\begin{figure}
\centering
\includegraphics[width=\columnwidth]{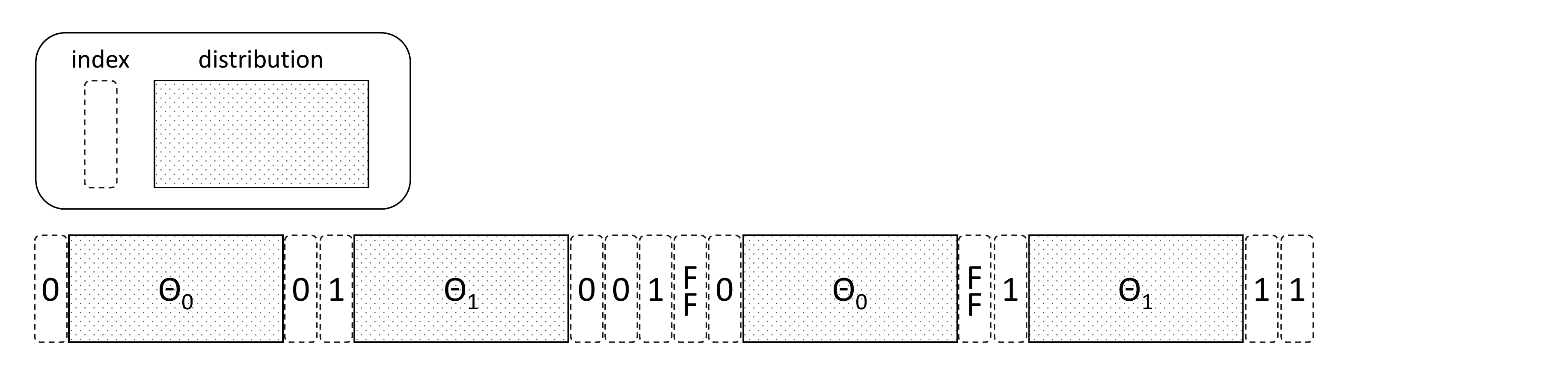}
\caption{An example of standard mode encoded stream structure by IDEALEM with $D=2$ and $B=16$ for time-series data shown in Fig.~\ref{fig_standard}. A dotted box represents an index in 1~byte; a solid box with pattern represents a source distribution $\Theta_j$ ($j=0,1$) in $8B$~bytes (i.e., 128~bytes). Note that \texttt{0xFF} denotes a special marker for overwriting signal.}
\label{fig_eight}
\end{figure}

Fig.~\ref{fig_standard} shows an example of random fluctuations of data where three groups of random variables exist. Here, each group shares the same underlying source distribution. If we assume two dictionary blocks (i.e., $D=2$) on IDEALEM,\footnote{Note that the number of dictionary blocks here is smaller than usual to demonstrate an encoded stream structure and buffer behavior. In most cases, it is recommended to use the maximum number $D=255$ for a higher compression ratio.} the time-series data in Fig.~\ref{fig_standard} can be encoded as shown in Fig.~\ref{fig_eight}.

In Fig.~\ref{fig_eight}, the first data block is outputted to an encoded stream \emph{as is}, along with the corresponding index that precedes the data block.\footnote{Counting starts from 0.} This data block is also stored in the buffer as the source distribution $\Theta_0$, which takes up $8B$ bytes in memory, as IDEALEM handles data in IEEE 754 double precision floating-point format where each sample is 8~bytes long. Since $B=16$ in Fig.~\ref{fig_standard}, each dictionary block occupies 128~bytes in this example.

The second data block is compared against the first dictionary block, and found to be exchangeable; thus the index~0 is solely outputted whose size is 1~byte. Then the third data block is encountered and compared against the first dictionary block which is the only block we have thus far; but it is not exchangeable and should be written on the encoded stream as well as the corresponding index. The data block is also stored in the buffer as $\Theta_1$. Now our buffer is full, as $D=2$.

The fourth and the fifth blocks are exchangeable with $\Theta_0$. Thus only the index~0 is outputted twice. The sixth block is first compared with the first dictionary entry ($\Theta_0$), but not exchangeable. It is next compared with the second entry ($\Theta_1$), and found to be exchangeable. So the index~1 is outputted to the encoded stream.

The seventh block is not exchangeable with any of two stored source probability distributions. Therefore, this data block should be stored in the buffer, which is impossible since all the dictionary entries are already occupied. IDEALEM currently discards the oldest dictionary entry in first-in-first-out (FIFO) manner, and replaces $\Theta_0$ with this data block. This overwriting should be signaled on the encoded stream so that the decoder can recognize it. To this end, IDEALEM uses a special marker \texttt{0xFF}, which automatically limits the number of dictionary blocks $D$ to the maximum of 255. This marker is first written on the encoded stream, and then the index and the data block are outputted as before.

The eighth block is not exchangeable either with any of two stored source distributions. In fact, it is exchangeable with one of previous data blocks (e.g., the first block) and would be represented with an index if we had more than two dictionary blocks (i.e., $D\geq3$), which means more compression. As discussed in Section~\ref{sec:operation}, more dictionary blocks in general promise higher compression ratios since there is a higher chance of finding a similar distribution stored in the buffer.

Thus the second dictionary entry ($\Theta_1$) should be replaced with the eighth data block and once again \texttt{0xFF} should be written on the encoded stream along with the index~1 and the data block. Finally, the ninth and the tenth blocks are exchangeable with $\Theta_1$ and the index~1 is outputted twice.

\subsubsection{Single Dictionary Block Case}
When the buffer can only hold a single dictionary entry, i.e., $D=1$, spending 1~byte on the index would waste the length of encoded stream, as an index would be always 0. Furthermore, frequent overwriting would lead to numerous \texttt{0xFF}'s each of which takes up 1~byte as well. Therefore, IDEALEM handles this as a special case.

\begin{figure}
\centering
\includegraphics[width=\columnwidth]{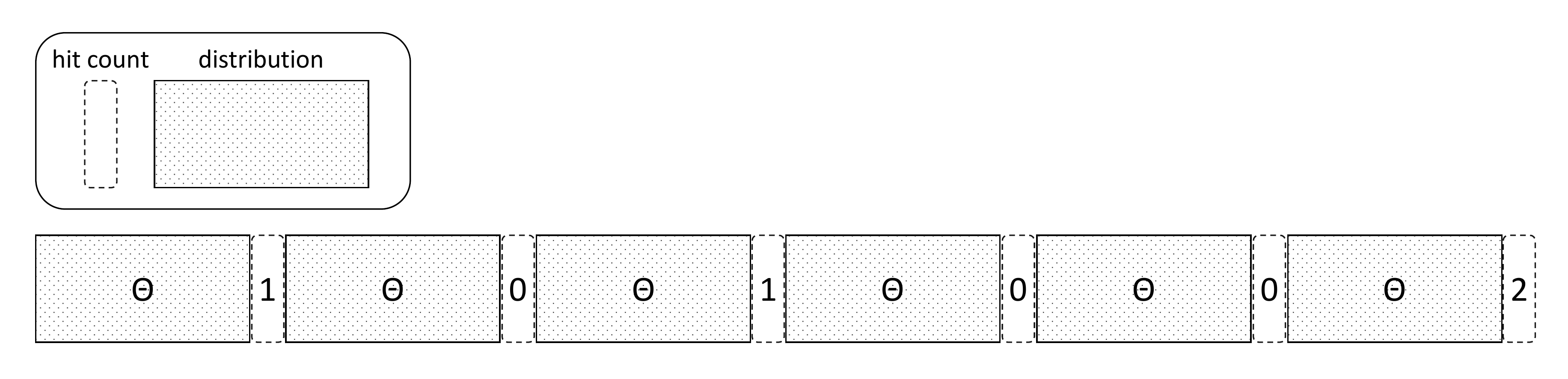}
\caption{An example of standard mode encoded stream structure when $D=1$ for time-series data shown in Fig.~\ref{fig_standard}. A dotted box represents a hit count in 1~byte; a solid box with pattern represents a source distribution $\Theta$ in $8B$~bytes (i.e., 128~bytes).}
\label{fig_nine}
\end{figure}

Fig.~\ref{fig_nine} shows an example of encoded stream structure in this case of $D=1$, which is the same scenario presented in Fig.~\ref{fig_eight}. However, the stream structure shown in Fig.~\ref{fig_nine} is different from the structure in Fig.~\ref{fig_eight}: indices are now replaced by \textbf{hit counts} and the positions of the solid box and the dotted box are exchanged. Since there is only one dictionary entry, a hit count records how many consecutive blocks are exchangeable with the previous source distribution.\footnote{Thus, a hit count 0 denotes that there is no repetition after a specific data block.}

In Fig.~\ref{fig_nine}, the second and the fifth blocks are exchangeable with their previous source distributions. Therefore, the hit counts are 1 for these data blocks. And the ninth and the tenth blocks are successively exchangeable with the previous source distribution. Thus the hit count 2 is written on the encoded stream for these data blocks. If we compare Fig.~\ref{fig_nine} with Fig.~\ref{fig_eight}, six source distributions are stored in Fig.~\ref{fig_nine}, while four source distributions are stored in Fig.~\ref{fig_eight}. And only three source distributions would be required if we had more than two dictionary blocks.

In the single dictionary block case, each hit count occupies 1~byte, allowing up to 255 repetitions to be represented by a single hit count.\footnote{More than 255 repetitions can be represented with another hit count, and so on.} However, while the hit count increases, the encoder cannot release it to the encoded stream, because the encoder has not encountered an unexchangeable data block yet. This phenomenon could happen especially when an input stream is monotonous, which would indefinitely delay the decoding process in an online streaming environment. In order to prevent this, IDEALEM has a \textbf{maximum count} parameter that controls decoding latency in the single dictionary block case. A smaller maximum count would enable a faster operation of IDEALEM in the streaming environment. On the other hand, it could also lengthen the encoded stream by using extra bytes for many repetitions.

\subsubsection{Decoding}
The encoded stream structures explained with Fig.~\ref{fig_eight} and Fig.~\ref{fig_nine} are in turn inputs to the decoder of IDEALEM. The decoder reconstructs time-series data from learned source distributions during the encoding process. This is accomplished with source probability distributions $\Theta_j$ and corresponding indices $j$ in the encoded stream; and in the single dictionary block case, with $\Theta$'s and corresponding hit counts. In Fig.~\ref{fig_eight} and Fig.~\ref{fig_nine}, data blocks that \emph{initiate} new source distributions are written on the encoded streams as is. Therefore, even though we could generate new data sequences out of these distributions, it is better to retain original data sequences for these initiating sequences during reconstruction.

On the other hand, it is impossible to reconstruct the same data sequence as the original for an exchangeable data block after the initiating sequence is reconstructed. Thus a new data sequence should be generated from a learned source distribution for this case, which means the order of data sequence can no longer be preserved. Since a stored distribution $\Theta_j$ or $\Theta$ is non-parametric, a random number generation from this distribution is equivalent to taking a random \emph{uniform} sample from stored data samples. This sampling is done \emph{without replacement} to avoid choosing any data sample more than once, which can be fundamentally regarded as a random permutation.

Another possibility of reconstructing data for the exchangeable data block would be simply duplicating a stored distribution for a new data sequence of exchangeable data block. However, this might generate artificial patterns in reconstructed data, which were not present in the original time-series data. With the random permutation, we can avoid any artificial patterns generated during the reconstruction process, which is studied using spectral analysis in Section~\ref{sec:duplication} and Section~\ref{sec:spectral}.

\subsection{Residual/Delta Mode} \label{sec:encoding_residual}
\begin{figure}
\centering
\includegraphics[width=\columnwidth]{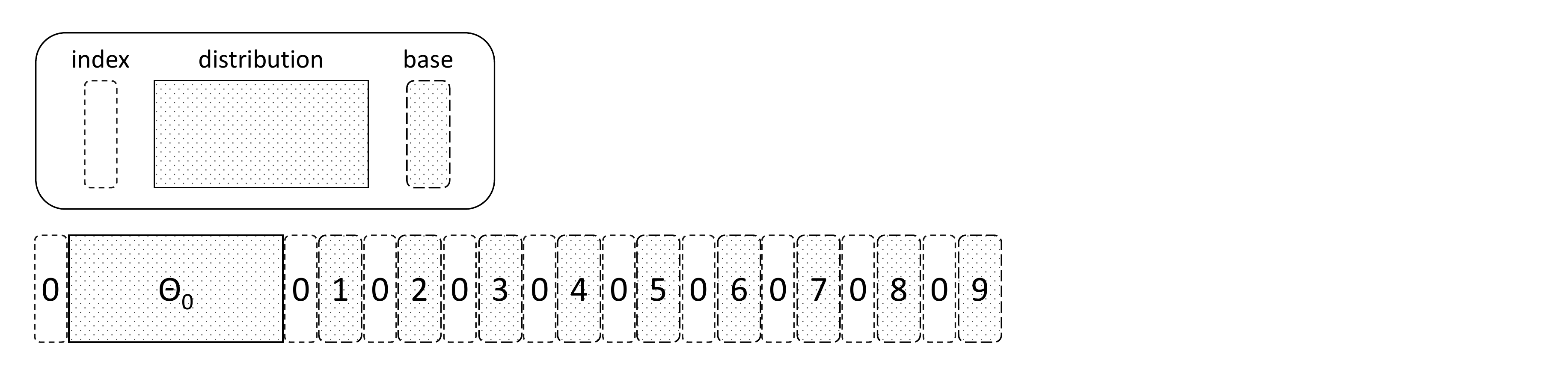}
\caption{An example of residual/delta mode encoded stream structure by IDEALEM with $D\geq2$ and $B=64$ for constantly increasing data shown in Fig.~\ref{fig_five}. A dotted box represents an index in 1~byte; a solid box with pattern represents the combination of a base value and a source distribution $\Theta_0$ in $8B$~bytes (i.e., 512~bytes); a dashed box with pattern represents a base value that is the first value of each data block.}
\label{fig_ten}
\end{figure}

Fig.~\ref{fig_five} shows an example of non-stationary data where every random variable, after transformation, is exchangeable with each other, which can be identified in Fig.~\ref{fig_five_b} and Fig.~\ref{fig_five_c} where residual/delta values look similar to each other in distributional sense. If we assume multiple dictionary blocks on IDEALEM, the data in Fig.~\ref{fig_five} can be encoded as shown in Fig.~\ref{fig_ten}.

Since only one dictionary entry is utilized in this example, the overwriting of dictionary block does not occur and thus the special marker \texttt{0xFF} is not shown in Fig.~\ref{fig_ten}. A big difference between the residual/delta mode and the standard mode is that in the residual/delta mode we have to write base values on the encoded stream along with indices and source distributions. Another difference is that a source distribution is composed of a base value and residual/delta values.

For instance, the first data block is outputted to the encoded stream after the transformation, along with the corresponding index before the data block. According to a specific transformation, the base value and residual or delta values are written on the encoded stream, which are also stored as the base value (8~bytes) and the source distribution $\Theta_0$ ($8(B-1)$~bytes) in the buffer. Note that it is unnecessary to store the base value itself in the buffer, because we only compare $B-1$ residual/delta values in the LEM processing. However, it is stored regardless to be compatible with the standard mode buffer operation.

All the other data blocks after the first data block are found to be exchangeable with the first dictionary entry ($\Theta_0$), when they are transformed. Thus the index~0, whose size is 1~byte, is outputted for each data block. In addition, a base value for each data block, whose size is 8~bytes, is separately written on the encoded stream after the index.

\subsubsection{Single Dictionary Block Case}
\begin{figure}
\centering
\includegraphics[width=\columnwidth]{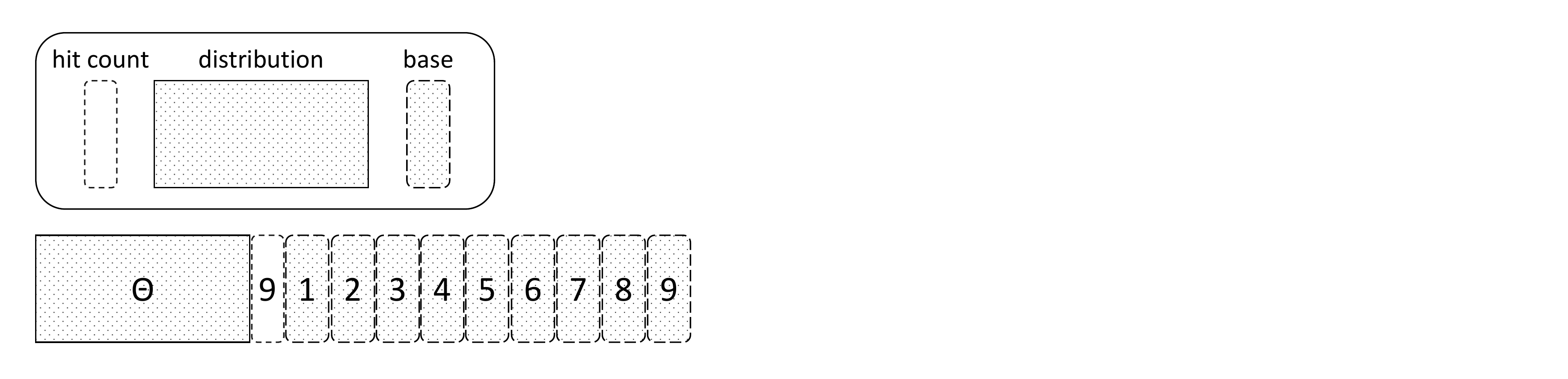}
\caption{An example of residual/delta mode encoded stream structure in the single dictionary block case ($D=1$) for the data shown in Fig.~\ref{fig_five}. A dotted box represents a hit count in 1~byte; a solid box with pattern represents the combination of a base value and a source distribution $\Theta$ in $8B$~bytes (i.e., 512~bytes); a dashed box with pattern represents a base value that is the first value of each data block.}
\label{fig_eleven}
\end{figure}

As in the standard mode, IDEALEM specially treats the single buffer case to save the length of encoded stream. Fig.~\ref{fig_eleven} shows the example of encoded stream structure in the single dictionary block case of the residual/delta mode, which is the same scenario presented in Fig.~\ref{fig_ten}. Since all data blocks after the first data block are exchangeable with the first dictionary entry ($\Theta$) that is composed of the base value and the residual/delta values,\footnote{Technically, base values do not play any role in the LEM processing.} there are nine hit counts in total. In addition, the base values for data blocks should be written on the encoded stream.

In Fig.~\ref{fig_eleven}, an interesting fact is that the single dictionary block case here can further reduce the length of encoded stream, as compared with the multiple dictionary blocks case in Fig.~\ref{fig_ten}. This can be attributed to the monotonicity of values after the residual/delta transformation in Fig.~\ref{fig_five}. In theory, it can be shown that the single dictionary block cases for both standard and residual/delta modes can achieve a higher compression ratio provided that data blocks to be compressed are so simple that most (if not all) of them can be represented with the previous distribution alone, which is further discussed in Section~\ref{sec:fundamental}. However, in reality, this condition could be difficult to fulfill because a single dictionary entry should be able to represent every possible source distribution.

\subsubsection{Decoding}
Again, the encoded stream structures in Fig.~\ref{fig_ten} and Fig.~\ref{fig_eleven} are inputs to the decoder of IDEALEM. The decoder reconstructs non-stationary data from learned residual/delta distributions and base values. This is accomplished with source residual/delta distributions $\Theta_j$, corresponding indices $j$, and corresponding base values in the encoded stream; and in the single dictionary block case, with $\Theta$'s, corresponding hit counts, and corresponding base values.

In contrast to the standard mode, decoding in the residual/delta mode does not perform random permutations from stored distributions when we handle the reconstruction of exchangeable data blocks after the initiating sequence is reconstructed. That is to say, IDEALEM simply adds a corresponding base value to residual/delta values (without the random permutation) for a new exchangeable data sequence. The random permutation should be avoided especially for the residual values because it would disorder values of increasing trend, which can be deduced from Fig.~\ref{fig_five_b}. Since we handle non-stationary data, which especially has constantly increasing trend, artificial patterns are not observable even without the random permutation.

\section{Theoretical Analysis} \label{sec:theoretical}
The encoded stream structures of two compression modes discussed in Section~\ref{sec:encoding} set theoretical upper bounds on achievable compression ratios. A notable difference between the standard mode and the residual/delta mode is the usage of the base value in the residual/delta mode, which decreases the compression ratio by the factor of nine. For the residual/delta mode, each transformation method shows different characteristics when processed by KS~test. In addition, the random permutation employed in the standard mode decoding affects the quality of reconstructed data.

\subsection{Fundamental Limit on Compression Performance} \label{sec:fundamental}
\subsubsection{Standard Mode} \label{sec:fundamental_standard}
With the encoded stream structure discussed in Section~\ref{sec:encoding_standard},
we inevitably set the maximum compression ratio we can achieve with a
given block size $B$. We can show this theoretical upper bound by the
following proposition.

\begin{proposition}
\label{pro:1}
Given a block size $B$, the maximum achievable compression ratio of
IDEALEM encoder with multiple dictionary blocks is $8\cdot B$.
\end{proposition}
\begin{IEEEproof}
  Without loss of generality, suppose we are compressing simple
  streaming data all of which can be represented with a single
  distribution $\Theta$. In other words, there is a single source of
  distribution that governs the generation of data in this random
  process. Then ideally we can represent the entire data stream (except
  the beginning part that composes $\Theta$) with many repetitions of
  the same index each of which takes up 1~byte.

  Assuming there are $i$ such repetitions, the original data size (in
  bytes) can be represented by $8B+8Bi$, where $8B$ is the size of the
  beginning part for $\Theta$; $8Bi$ is the size of the entire data
  stream excluding the beginning part. On the other hand, the compressed
  data size is represented by $1+8B+i$, where $1$ and $8B$ are the sizes
  of the initial index and $\Theta$; $i$ is the size of repeating
  indices. If continuous streaming of data is assumed, the
  compression ratio is given by
\begin{equation}
\label{eq_pro1}
\lim_{i\to\infty}\frac{8B+8Bi}{1+8B+i}=8\cdot B.
\end{equation}

The compression ratio (\ref{eq_pro1}) is the maximum achievable
compression ratio, because in a scenario where the exchangeability is
not guaranteed, we would eventually have frequent overwriting of
$\Theta$, which no longer allows us to use the constant term $8B$ and
$1+8B$ in (\ref{eq_pro1}).

A similar claim can be also made in the case where streaming data is
composed of multiple \emph{bounded} sources of distributions. Since this
can be covered with source distributions $\Theta_j$ and corresponding indices $j$ in the
encoded stream, we can use a bounded number of constant terms for the
compression ratio, which would again result in $8\cdot B$.
\end{IEEEproof}

\begin{corollary}
\label{cor:1}
For the single dictionary block case, the maximum achievable compression ratio of
IDEALEM encoder with a maximum count $c$ is $8\cdot cB$.
\end{corollary}
\begin{IEEEproof}
  Again, suppose we are compressing simple streaming data all of which
  can be represented with a single distribution $\Theta$. Then we can
  represent the entire data stream (except the beginning part that
  composes $\Theta$) with many hit counts each of which can record up to
  $c$ repetitions.

  Assuming there are $i$ such repetitions, the original data size can
  again be represented by $8B+8Bi$. On the other hand, the compressed
  data size is represented by $8B+\lceil{i/c}\rceil$, where $8B$ is the
  size of $\Theta$; $\lceil{i/c}\rceil$ is the size of hit
  counts. Assuming continuous streaming of data, the compression ratio
  is given by
\begin{equation}
\label{eq_cor1}
\lim_{i\to\infty}\frac{8B+8Bi}{8B+\lceil{i/c}\rceil}=8\cdot cB.
\end{equation}

The compression ratio (\ref{eq_cor1}) is the maximum compression ratio
and can be achievable if and only if there is a bounded number of
distribution changes (including no change) in streaming data.
\end{IEEEproof}

Proposition~\ref{pro:1} indicates that a large $B$ potentially increases
compression ratio. However, a large $B$ also increases the difficulty of
passing the KS~test due to the sensitivity discussed in
Section~\ref{sec:similarity}. In practice, we cannot have ideal
streaming data whose data sequences are nearly identical in terms of the maximum
distributional distance (\ref{eq_one}). Thus it is difficult to achieve
the compression ratio of $8\cdot B$ in real-world data, when $B$ is large.

Corollary~\ref{cor:1} pushes this compression ratio even further: it can
be as high as $2040\cdot B$ with $c=255$. This condition is obviously
more difficult to achieve in reality, as we have to check the
exchangeability with the previous distribution alone.

\subsubsection{Residual/Delta Mode}
For the residual/delta mode whose operation is discussed in Section~\ref{sec:encoding_residual}, the encoded stream structures shown in Fig.~\ref{fig_ten} and Fig.~\ref{fig_eleven} set the theoretical upper bounds of achievable compression ratios with the following proposition and corollary.

\begin{proposition}
\label{pro:2}
Given a block length $B$, the maximum achievable compression ratio of
residual/delta mode in IDEALEM with multiple dictionary blocks is $(8/9)\cdot B$.
\end{proposition}
\begin{IEEEproof}
Without loss of generality, suppose we are compressing monotonically changing data where every random variable, after transformation, can be represented with a single distribution $\Theta$. In other words, there is a single source of distribution that governs the generation of residual/delta values in this random process. Then we can represent the entire data stream (except the beginning part that contains $\Theta$) with many repetitions of the same index (1~byte) and corresponding base values (8~bytes).

Assuming there are $i$ such repetitions, the original data size (in bytes) can be represented by $8B+8Bi$, where $8B$ is the size of the beginning part that corresponds to $\Theta$; $8Bi$ is the size of the entire data stream excluding the beginning part. On the other hand, the compressed data size is represented by $1+8+8(B-1)+(1+8)i$, where $1$, $8$, and $8(B-1)$ are the sizes of the initial index, the base value, and $\Theta$; $(1+8)i$ is the size of repeating indices and corresponding base values. If continuous streaming of data is assumed, the compression ratio is given by
\begin{equation}
\label{eq_pro2}
\lim_{i\to\infty}\frac{8B+8Bi}{1+8B+9i}=(8/9)\cdot B.
\end{equation}

The compression ratio (\ref{eq_pro2}) is the maximum achievable compression ratio, because in a scenario where the exchangeability is not guaranteed, we would eventually have frequent overwriting of $\Theta$, which no longer allows us to use the constant term $8B$ and $1+8B$ in (\ref{eq_pro2}).

A similar claim can be also made in the case where monotonically changing data is composed of multiple \emph{bounded} sources of distributions that account for residual/delta values. Since this can be covered with $\Theta_j$'s and corresponding indices $j$ and base values in the encoded stream, we can use a bounded number of constant terms for the compression ratio, which would again result in $(8/9)\cdot B$.
\end{IEEEproof}

\begin{corollary}
\label{cor:2}
For the single dictionary block case, the maximum achievable compression ratio of
residual/delta mode in IDEALEM with a maximum count $c$ is $8cB/(1+8c)$.
\end{corollary}
\begin{IEEEproof}
Again, suppose we are compressing monotonically changing data where every random variable, after transformation, can be represented with a single distribution $\Theta$. Then we can represent the entire data stream (except the beginning part that composes $\Theta$) with many hit counts each of which can record up to $c$ repetitions and base values that correspond to the hit counts.

Assuming there are $i$ such repetitions, the original data size can again be represented by $8B+8Bi$. On the other hand, the compressed data size is represented by $8+8(B-1)+\lceil{i/c}\rceil+8i$, where $8$ and $8(B-1)$ are the sizes of the base value and $\Theta$; $\lceil{i/c}\rceil$ is the size of hit counts; $8i$ is the size of base values each of which is needed for each hit. Assuming again continuous streaming of data, the compression ratio is given by
\begin{equation}
\label{eq_cor2}
\lim_{i\to\infty}\frac{8B+8Bi}{8B+\lceil{i/c}\rceil+8i}=8cB/(1+8c).
\end{equation}

The compression ratio (\ref{eq_cor2}) is the maximum compression ratio and can be achievable if and only if there is a bounded number of distribution changes (including no change) for residual/delta values in monotonically changing data.
\end{IEEEproof}

In the standard mode, the maximum achievable compression ratio with multiple dictionary blocks is $8\cdot B$; whereas Proposition~\ref{pro:2} tells that it is $(8/9)\cdot B$ in the residual/delta mode. The ninefold decrease in the compression ratio for the residual/delta mode is due to the necessity of base values: in a nutshell, the standard mode uses only 1~byte to represent a data block in the encoded stream, but the residual/delta mode uses 9~bytes to represent a data block on the other hand.

Similar to Proposition~\ref{pro:1}, Proposition~\ref{pro:2} also indicates that a large $B$ potentially increases the compression ratio of residual/delta encoding, although it also increases the difficulty of passing the KS~test. Increasing $B$ does not necessarily yield a higher compression ratio as discussed in Section~\ref{sec:fundamental_standard}.

The compression ratio with a single dictionary block shown in Corollary~\ref{cor:2} is a small improvement over the compression ratio with multiple dictionary blocks. In fact, it becomes equivalent to $(8/9)\cdot B$ with $c=1$, which is the maximum achievable compression ratio with multiple dictionary blocks. Although it can increase theoretically as high as $(2040/2041)\cdot B$, it is still far from $8\cdot cB$ we can achieve with the standard mode shown in Corollary~\ref{cor:1}. This is again attributable to the necessity of base values in the residual/delta mode: each base value takes up 8~bytes and the size of base values dominates the size of the encoded stream structure.

As in Proposition~\ref{pro:2}, a large $B$ could increase the compression ratio with the single buffer case; but it also increases the difficulty of passing the KS~test at the same time. Furthermore, checking the exchangeability with the previous distribution alone makes it difficult to achieve the fundamental limit of the compression ratio with a single dictionary block.

\subsection{Exchangeability with Residual and Delta Transformations}
Fig.~\ref{fig_five} shows the difference between the patterns of residual and delta transformation methods: the residual transformation still retains the increasing trend in data, whereas the delta transformation removes the trend and makes resulting values \emph{uncorrelated} with each other. With this observation, we can analyze how it affects the LEM processing by KS~test.

Without loss of generality, we consider a monotonically changing (increasing or decreasing) time series as follows ($i=0,1,2,\ldots$):
\begin{equation}
\label{eq_eight}
x^*_{i+1}=x^*_i+m,
\end{equation}
where $x^*_i$ denotes a noiseless value and $m$ a constant that represents increment or decrement. In reality, each value in time series cannot be free from noise due to many factors such as sensor and device precision. To take this into account, a new time series model with noise is given by
\begin{equation}
\label{eq_nine}
x_i=x^*_i+w_i,
\end{equation}
where $w_i$ is independent and identically distributed (i.i.d.) variable across different $i$'s. In particular, we assume that $w_i$ is zero-mean Gaussian random variable with variance $\sigma^2_w$ such that $w_i\sim\mathcal{N}(0,\sigma^2_w)$.

As discussed in Section~\ref{sec:transformations}, we use $x^r_{jB+k}$ in (\ref{eq_four}) and $x^d_{jB+k}$ in (\ref{eq_six}) for the residual transformation and the delta transformation, respectively. We can then plug our time series model into (\ref{eq_four}) and (\ref{eq_six}) to gain insight into differences between residual and delta transformations. For the residual transformation, we have
\begin{equation}
\label{eq_ten}
x_{jB+k}-x_{jB}=x^*_{jB+k}+w_{jB+k}-x^*_{jB}-w_{jB}=km+w_{jB+k}-w_{jB}.
\end{equation}
From (\ref{eq_ten}), we can see $x^r_{jB+k}\sim\mathcal{N}(km,2\sigma^2_w)$ ($k=1,2,\ldots,B-1$) using the properties of Gaussian distribution. Similarly for the delta transformation, we have
\begin{equation}
\label{eq_eleven}
x_{jB+k}-x_{jB+k-1}=x^*_{jB+k}+w_{jB+k}-x^*_{jB+k-1}-w_{jB+k-1}=m+w_{jB+k}-w_{jB+k-1}.
\end{equation}
From (\ref{eq_eleven}), we see $x^d_{jB+k}\sim\mathcal{N}(m,2\sigma^2_w)$ again using the properties of Gaussian distribution, where we can observe that $x^d_{jB+k}$ no longer depends on $k$. This explains why the delta transformation makes resulting values uncorrelated with each other, while the residual transformation retains the increasing trend in data.

It should be noted that the LEM processing is performed on the block level: $b^r_j$ for the residual transformation or $b^d_j$ for the delta transformation. In particular, an empirical distribution of each block is determined by the group of $B-1$ values $x^r_{jB+k}$ or $x^d_{jB+k}$. Thus we can model its distribution on the \emph{mixture} of Gaussians. In particular, we have
\begin{equation}
\label{eq_add1}
b^r_j\sim\sum_{k=1}^{B-1}\frac{1}{B-1}\mathcal{N}(km,2\sigma^2_w)
\end{equation}
for the residual transformation, where the distribution is represented by the combination of $x^r_{jB+k}$'s. For the delta transformation, we have
\begin{equation}
\label{eq_add2}
b^d_j\sim\sum_{k=1}^{B-1}\frac{1}{B-1}\mathcal{N}(m,2\sigma^2_w)=\mathcal{N}(m,2\sigma^2_w).
\end{equation}
In (\ref{eq_add2}), we notice that the distribution for $b^d_j$ is eventually the same as $x^d_{jB+k}$, due to the independence of $x^d_{jB+k}$ from $k$.

\subsubsection{Similar Block Model} \label{sec:model}
The notion of similarity entails small difference: data blocks similar to each other also have small differences. To model these differences, we devise two similar block cases that are also mathematically tractable: a similar block of which all values are slightly different from the values of the time series model (\ref{eq_nine}); a similar block of which some values are quite off from the values of the time series model (\ref{eq_nine}).

\begin{definition} \label{def:1}
A similar block of the first kind is generated by adding another white noise to existing time series, which is defined by
\begin{equation}
\label{eq_def1}
x'_i=x_i+w'_i,
\end{equation}
where $w'_i$ is i.i.d. across different $i$'s and $w'_i\sim\mathcal{N}(0,\sigma^2_{w'})$.
\end{definition}

Using $x'_i$ and the properties of Gaussian distribution, we can compute the residual transformation $x^r_{jB+k}$ as follows:
\begin{equation}
\label{eq_twelve}
x'_{jB+k}-x'_{jB}=x^*_{jB+k}+w_{jB+k}+w'_{jB+k}-x^*_{jB}-w_{jB}-w'_{jB}=km+w_{jB+k}+w'_{jB+k}-w_{jB}-w'_{jB}.
\end{equation}
Therefore, in the case of the similar block of the first kind, we observe $x^r_{jB+k}\sim\mathcal{N}(km,2(\sigma^2_w+\sigma^2_{w'}))$ ($k=1,2,\ldots,B-1$). Similarly for the delta transformation $x^d_{jB+k}$, we have
\setlength{\arraycolsep}{0.0em}
\begin{eqnarray}
\label{eq_thirteen}
x'_{jB+k}-x'_{jB+k-1}&{}={}&x^*_{jB+k}+w_{jB+k}+w'_{jB+k}-x^*_{jB+k-1}-w_{jB+k-1}-w'_{jB+k-1}\nonumber\\
 &{}={}&m+w_{jB+k}+w'_{jB+k}-w_{jB+k-1}-w'_{jB+k-1}.
\end{eqnarray}
\setlength{\arraycolsep}{5pt}%
Thus, in the case of the similar block of the first kind, we observe $x^d_{jB+k}\sim\mathcal{N}(m,2(\sigma^2_w+\sigma^2_{w'}))$.

Since the distributions of $x^r_{jB+k}$ and $x^d_{jB+k}$ are given, we can consider the block-level distributions using the mixture of Gaussians as follows:
\begin{equation}
\label{eq_add3}
b^r_j\sim\sum_{k=1}^{B-1}\frac{1}{B-1}\mathcal{N}(km,2(\sigma^2_w+\sigma^2_{w'}))
\end{equation}
for the residual transformation; and
\begin{equation}
\label{eq_add4}
b^d_j\sim\sum_{k=1}^{B-1}\frac{1}{B-1}\mathcal{N}(m,2(\sigma^2_w+\sigma^2_{w'}))=\mathcal{N}(m,2(\sigma^2_w+\sigma^2_{w'}))
\end{equation}
for the delta transformation. Note that these distributions for the similar blocks of the first kind are close to (\ref{eq_add1}) and (\ref{eq_add2}), except for the additional white noise variance $2\sigma^2_{w'}$.

\begin{definition} \label{def:2}
A similar block of the second kind is generated by dividing a block $j$ into two groups with a fraction $\alpha$\footnote{Without loss of generality, we assume that $\alpha$ is restricted such that $\alpha B\in\mathbb{Z}$.}: the first group identical to existing time series and the second group having the same values, which is defined by
\begin{equation}
\label{eq_def2}
x''_{jB+k}=\left\{\begin{array}{cc}
x_{jB+k} & k=0,1,\ldots,\alpha B\\
x_{jB+\alpha B} & k=\alpha B+1,\alpha B+2,\ldots,B-1
\end{array} \right..
\end{equation}
\end{definition}

With this new definition for the similar block of the second kind, we can compute the residual transformation $x^r_{jB+k}$ for two groups. For the first group, we have $x''_{jB+k}-x''_{jB}=x_{jB+k}-x_{jB}$, which is identical to (\ref{eq_ten}). Therefore $x^r_{jB+k}\sim\mathcal{N}(km,2\sigma^2_w)$ ($k=1,2,\ldots,\alpha B$). As for the second group ($k=\alpha B+1,\alpha B+2,\ldots,B-1$), it is clear that $x^r_{jB+k}\sim\mathcal{N}(\alpha Bm,2\sigma^2_w)$ because $x^r_{jB+k}=x^r_{jB+\alpha B}$.

Next, we compute the delta transformation $x^d_{jB+k}$ for two groups. For the first group ($k=1,2,\ldots,\alpha B$), we have $x''_{jB+k}-x''_{jB+k-1}=x_{jB+k}-x_{jB+k-1}$, which is identical to (\ref{eq_eleven}). Therefore $x^d_{jB+k}\sim\mathcal{N}(m,2\sigma^2_w)$. For the second group ($k=\alpha B+1,\alpha B+2,\ldots,B-1$), we have
\begin{equation}
\label{eq_fourteen}
x''_{jB+k}-x''_{jB+k-1}=x_{jB+\alpha B}-x_{jB+\alpha B}=0.
\end{equation}
Thus, $x^d_{jB+k}=0$ ($k=\alpha B+1,\alpha B+2,\ldots,B-1$). Note that $x^d_{jB+k}$ is no longer stochastic for the second group.

We now consider the block-level distributions. In particular, the block distribution of the residual transformation is given by
\setlength{\arraycolsep}{0.0em}
\begin{eqnarray}
\label{eq_fifteen}
b^r_j&{}\sim{}&\sum_{k=1}^{\alpha B}\frac{1}{B-1}\mathcal{N}(km,2\sigma^2_w)+\sum_{k=\alpha B+1}^{B-1}\frac{1}{B-1}\mathcal{N}(\alpha Bm,2\sigma^2_w)\nonumber\\
 &{}={}&\sum_{k=1}^{\alpha B-1}\frac{1}{B-1}\mathcal{N}(km,2\sigma^2_w)+\frac{B-\alpha B}{B-1}\mathcal{N}(\alpha Bm,2\sigma^2_w).
\end{eqnarray}
For the block distribution of the delta transformation, we use the mixture of Gaussian and the \emph{Dirac delta function} as follows:
\begin{eqnarray}
\label{eq_sixteen}
b^d_j&{}\sim{}&\sum_{k=1}^{\alpha B}\frac{1}{B-1}\mathcal{N}(m,2\sigma^2_w)+\sum_{k=\alpha B+1}^{B-1}\frac{1}{B-1}\delta(\cdot)\nonumber\\
 &{}={}&\frac{\alpha B}{B-1}\mathcal{N}(m,2\sigma^2_w)+\frac{B-1-\alpha B}{B-1}\delta(\cdot),
\end{eqnarray}
\setlength{\arraycolsep}{5pt}%
where $\delta(\cdot)$ indicates a probability mass at zero.

\subsubsection{Exchangeability Analysis}
By comparing the similar block model against the block comprised of the original time series model (\ref{eq_nine}), we can better understand difference in exchangeability between the residual and delta transformations. Specifically, we are interested to see which transformation makes data blocks more easily exchangeable for each definition in Section~\ref{sec:model}. We use block-level distributions and the maximum distributional distance between two random variables (\ref{eq_one}) to assess the exchangeability.

\begin{lemma}
\label{lem:1}
For the similar block of the first kind, the residual transformation allows more exchangeability than the delta transformation does.
\end{lemma}
\begin{IEEEproof}
We first compare two $b^d_j$'s (\ref{eq_add4}) and (\ref{eq_add2}) in terms of (\ref{eq_one}). Since the cumulative distribution function (CDF) of $\mathcal{N}(x\mid\mu,\sigma^2)$ is known to be $\Phi(\frac{x-\mu}{\sigma})=\frac{1}{2}(1+\erf(\frac{x-\mu}{\sqrt{2}\sigma}))$, we first write the distributional distance between (\ref{eq_add4}) and (\ref{eq_add2}) as follows:
\begin{equation}
\label{eq_seventeen}
\frac{1}{2}\left(1+\erf\left(\frac{x-m}{2\sqrt{\sigma^2_w+\sigma^2_{w'}}}\right)\right)-\frac{1}{2}\left(1+\erf\left(\frac{x-m}{2\sigma_w}\right)\right),
\end{equation}
where $\erf(\cdot)$ is the Gauss error function. Then we can find the maximum distributional distance (\ref{eq_one}) by setting the derivative of (\ref{eq_seventeen}) equal to zero, which is given by
\begin{equation}
\label{eq_eighteen}
\frac{1}{2\sqrt{\pi(\sigma^2_w+\sigma^2_{w'})}}\exp\left(-\left(\frac{x-m}{2\sqrt{\sigma^2_w+\sigma^2_{w'}}}\right)^2\right)-\frac{1}{2\sqrt{\pi}\sigma_w}\exp\left(-\left(\frac{x-m}{2\sigma_w}\right)^2\right)=0.
\end{equation}
Rewriting (\ref{eq_eighteen}) in terms of $x$, we have
\begin{equation}
\label{eq_nineteen}
x=m\pm\sqrt{\frac{4\sigma^2_w(\sigma^2_w+\sigma^2_{w'})}{\sigma^2_{w'}}\cdot\xi},
\end{equation}
where $\xi=\ln(\sqrt{\sigma^2_w+\sigma^2_{w'}}/\sigma_w)$.
Both solutions in (\ref{eq_nineteen}) yield the maximum distributional distance thanks to the absolute value in (\ref{eq_one}), which is given by
\begin{equation}
\label{eq_twenty}
\frac{1}{2}\erf\left(\frac{\sqrt{\sigma^2_w+\sigma^2_{w'}}\cdot\sqrt{\xi}}{\sigma_{w'}}\right)-\frac{1}{2}\erf\left(\frac{\sigma_w\sqrt{\xi}}{\sigma_{w'}}\right).
\end{equation}

Next we compare two $b^r_j$'s (\ref{eq_add3}) and (\ref{eq_add1}) and write the distributional distance between them in the form
\begin{equation}
\label{eq_twentyone}
\sum_{k=1}^{B-1}\frac{1}{2(B-1)}\left(\erf\left(\frac{x-km}{2\sqrt{\sigma^2_w+\sigma^2_{w'}}}\right)-\erf\left(\frac{x-km}{2\sigma_w}\right)\right).
\end{equation}
Since we have $B-1$ summands in (\ref{eq_twentyone}), we cannot simply utilize the same technique used for $b^d_j$ to find the maximum distributional distance. Specifically, (\ref{eq_twentyone}) has $2(B-1)$ maxima and minima, which makes it difficult to find analytic solutions.

On the other hand, it should be noted that $B-1$ Gaussian distributions for each $b^r_j$ only differ in means (locations); so they are all in the same shape. This means we can choose any of $2(B-1)$ maxima and minima for the maximum distributional distance, especially when the additional white noise variance $2\sigma^2_{w'}$ is small and the tails of $B-1$ Gaussian distributions in (\ref{eq_add3}) hardly overlap. However, if $2\sigma^2_{w'}$ gets larger and the tails start to overlap, valleys between peaks of $B-1$ Gaussian distributions get shallower. This leads to changes in the CDF of (\ref{eq_add3}); the absolute values of maxima and minima between two edges are smaller than those on the edges ($x<m$ and $x>(B-1)m$) in (\ref{eq_twentyone}). Thus we can assume that regardless of $2\sigma^2_{w'}$, the global maximum occurs when $x<m$ and the global minimum when $x>(B-1)m$ in (\ref{eq_twentyone}).

For the sake of simplicity, we focus on the global maximum case $x<m$. When $2\sigma^2_{w'}$ is small, we can ignore all summands but the first one in (\ref{eq_twentyone}), which becomes almost identical to (\ref{eq_seventeen}). Thus the maximum distributional distance is essentially (\ref{eq_twenty}) reduced by a factor of $B-1$. On the other hand, when $2\sigma^2_{w'}$ is large, we should take more summands into account and the maximum distributional distance would be larger than the case when $2\sigma^2_{w'}$ is small. Nonetheless, we have the following relationship for $x<m$:
\setlength{\arraycolsep}{0.0em}
\begin{eqnarray}
\label{eq_twentytwo}
\max\left(\erf\left(\frac{x-m}{2\sqrt{\sigma^2_w+\sigma^2_{w'}}}\right)-\erf\left(\frac{x-m}{2\sigma_w}\right)\right)>&&\nonumber\\
\max\left(\erf\left(\frac{x-2m}{2\sqrt{\sigma^2_w+\sigma^2_{w'}}}\right)-\erf\left(\frac{x-2m}{2\sigma_w}\right)\right)>&{}\cdots{}&>\nonumber\\
\max\left(\erf\left(\frac{x-(B-1)m}{2\sqrt{\sigma^2_w+\sigma^2_{w'}}}\right)-\erf\left(\frac{x-(B-1)m}{2\sigma_w}\right)\right).&&
\end{eqnarray}
\setlength{\arraycolsep}{5pt}%

Using (\ref{eq_twentytwo}), we note that the global maximum of (\ref{eq_twentyone}) is always less than (\ref{eq_twenty}) for any $2\sigma^2_{w'}$. This is because the global maximum should be less than the summation of $B-1$ maxima of each summand in (\ref{eq_twentyone}). Therefore, we see that the distributional distance for $b^r_j$ is smaller than that for $b^d_j$, which implies more exchangeability for $b^r_j$.
\end{IEEEproof}

\begin{lemma}
\label{lem:2}
For the similar block of the second kind, the delta transformation allows more exchangeability than the residual transformation does.
\end{lemma}
\begin{IEEEproof}
Again, we first compare two $b^d_j$'s (\ref{eq_sixteen}) and (\ref{eq_add2}) in terms of (\ref{eq_one}). And we write the distributional distance between (\ref{eq_sixteen}) and (\ref{eq_add2}) as follows:
\begin{equation}
\label{eq_twentythree}
\frac{B-1-\alpha B}{B-1}H(x)-\frac{B-1-\alpha B}{B-1}\Phi\left(\frac{x-m}{\sqrt{2}\sigma_w}\right),
\end{equation}
where $H(\cdot)$ indicates the Heaviside step function. In (\ref{eq_twentythree}), the maximum distributional distance (\ref{eq_one}) is given when we set $x=0$, which is in the form
\begin{equation}
\label{eq_twentythreehalf}
\frac{B-1-\alpha B}{B-1}\left(1-\Phi\left(\frac{-m}{\sqrt{2}\sigma_w}\right)\right).
\end{equation}

Next we compare two $b^r_j$'s (\ref{eq_fifteen}) and (\ref{eq_add1}), whose distributional distance is given by
\begin{equation}
\label{eq_twentyfour}
\sum_{k=\alpha B+1}^{B-1}\frac{1}{B-1}\left(\Phi\left(\frac{x-\alpha Bm}{\sqrt{2}\sigma_w}\right)-\Phi\left(\frac{x-km}{\sqrt{2}\sigma_w}\right)\right).
\end{equation}
With the assumption of small $2\sigma^2_w$ such that the tails of $B-1$ Gaussian distributions in (\ref{eq_add1}) do not overlap, we note that $\Phi\left(\frac{x-km}{\sqrt{2}\sigma_w}\right)=0$ when $x<(k-0.5)m$; and $\Phi\left(\frac{x-km}{\sqrt{2}\sigma_w}\right)=1$ when $x>(k+0.5)m$. Using this property, we first consider two cases to find the maximum value of (\ref{eq_twentyfour}): $x<(\alpha B+0.5)m$ and $x>(\alpha B+0.5)m$.

For $x<(\alpha B+0.5)m$, (\ref{eq_twentyfour}) reduces to
\begin{equation}
\label{eq_twentyfive}
\frac{B-1-\alpha B}{B-1}\Phi\left(\frac{x-\alpha Bm}{\sqrt{2}\sigma_w}\right),
\end{equation}
whereas, for $x>(\alpha B+0.5)m$, we have
\begin{equation}
\label{eq_twentysix}
\sum_{k=\alpha B+1}^{B-1}\frac{1}{B-1}\left(1-\Phi\left(\frac{x-km}{\sqrt{2}\sigma_w}\right)\right).
\end{equation}
Considering both (\ref{eq_twentyfive}) and (\ref{eq_twentysix}), we observe that we can obtain the maximum value $\frac{B-1-\alpha B}{B-1}$ by setting $x=(\alpha B+0.5)m$, which is essentially the maximum distributional distance.

Since we have the following relationship
\begin{equation}
\label{eq_twentyseven}
\frac{B-1-\alpha B}{B-1}>\frac{B-1-\alpha B}{B-1}\left(1-\Phi\left(\frac{-m}{\sqrt{2}\sigma_w}\right)\right),
\end{equation}
the distributional distance for $b^d_j$ is smaller than that for $b^r_j$. This implies more exchangeability for $b^d_j$ and concludes the proof.
\end{IEEEproof}

\begin{theorem}
\label{thm:1}
Each transformation method for non-stationary data has a varying degree of exchangeability for different types of similar data sequences. Therefore one method shows better compression performance than the other method does and vice versa depending on data types.
\end{theorem}
\begin{IEEEproof}
Lemma~\ref{lem:1} shows that the residual transformation is more advantageous to exchangeability for the similar block of the first kind, whereas Lemma~\ref{lem:2} shows that the delta transformation is more advantageous for the similar block of the second kind. Since more exchangeability leads to better compression performance, the residual transformation may yield better compression performance for data sequences characterized by the similar block of the first kind; whereas the delta transformation for data sequences characterized by the similar block of the second kind.
\end{IEEEproof}
Theorem~\ref{thm:1} indicates that there is no clear winner in transformation methods from the perspective of compression performance, which will be demonstrated in Section~\ref{sec:comparison}.

\subsection{Effect of Duplication on Exchangeable Data Blocks} \label{sec:duplication}
Section~\ref{sec:encoding} discussed the strategy of reconstructing exchangeable data blocks during the decoding process after the initiating sequence was already reconstructed: the standard mode randomly permutes the order of data sequence from learned source distributions to avoid artificial and repetitive patterns, whereas the residual/delta mode forgoes the random permutation as artificial patterns are hardly observable in this mode thanks to the nature of residual/delta values.

To study the effect of skipping the random permutation (i.e., duplication) in the frequency domain, we perform a spectral analysis and show that duplicating data blocks has an effect on frequency components of reconstructed data, which is explained in the following proposition.

\begin{proposition}
\label{pro:3}
Duplicating an exchangeable data block $K$ times concentrates energy on $k$th frequency components where $k$ are multiples of $K$.
\end{proposition}
\begin{IEEEproof}
Suppose we have a data block with values $\Theta=(x_0, x_1, \ldots, x_{B-1})$. We can perform the discrete Fourier transform (DFT) on this data block by $\widetilde{F}_k=\sum_{n=0}^{B-1}x_n\exp(-j\frac{2\pi kn}{B})$, where $\widetilde{F}_k$ ($k=0,1,\ldots,B-1$) are frequency components of the original data block $\Theta$.

Now we assume there is a single source of distribution $\Theta$ and all other data blocks are exchangeable with $\Theta$. We can duplicate $\Theta$ for $K$ times (including the first occurrence) such that
\begin{equation}
\label{eq_pro3}
x_n=\left\{\begin{array}{cc}
x_n & n<B\\
x_{n-B} & B\leq n<2B\\
\vdots &\\
x_{n-(K-1)B} & (K-1)B\leq n
\end{array} \right.,
\end{equation}
where $n=0,1,\ldots,KB-1$.

We are interested in the DFT of this new data sequence as follows:
\setlength{\arraycolsep}{0.0em}
\begin{eqnarray}
\label{eq_pro4}
F_k&{}={}&\sum_{n=0}^{KB-1}x_n e^{-j\frac{2\pi kn}{KB}}\nonumber\\
 &{}={}&\sum_{n=0}^{B-1}x_n e^{-j\frac{2\pi kn}{KB}}+\sum_{n=B}^{2B-1}x_{n-B}e^{-j\frac{2\pi kn}{KB}}+\cdots+\sum_{n=(K-1)B}^{KB-1}x_{n-(K-1)B}e^{-j\frac{2\pi kn}{KB}}\nonumber\\
 &{}={}&\sum_{n=0}^{B-1}x_n\sum_{i=0}^{K-1}e^{-j\frac{2\pi k(iB+n)}{KB}}\nonumber\\
 &{}={}&\sum_{n=0}^{B-1}x_n\sum_{i=0}^{K-1}e^{-j\frac{2\pi kiB}{KB}}e^{-j\frac{2\pi kn}{KB}}\nonumber\\
 &{}={}&\sum_{n=0}^{B-1}x_n e^{-j\frac{2\pi kn}{KB}}\sum_{i=0}^{K-1}e^{-j\frac{2\pi ki}{K}}.
\end{eqnarray}
\setlength{\arraycolsep}{5pt}%
From (\ref{eq_pro4}), we see
\begin{equation}
\label{eq_pro5}
\sum_{i=0}^{K-1}e^{-j\frac{2\pi ki}{K}}=\left\{\begin{array}{cc}
K & k=0,K,2K,\ldots\\
0 & \textrm{otherwise}
\end{array} \right..
\end{equation}
Therefore, we can rewrite (\ref{eq_pro4}) in the following form
\begin{equation}
\label{eq_pro6}
F_k=\left\{\begin{array}{cc}
K\widetilde{F}_{k/K} & k=0,K,2K,\ldots\\
0 & \textrm{otherwise}
\end{array} \right.,
\end{equation}
which is represented with the frequency components of the original data block $\Theta$. We notice that (\ref{eq_pro6}) indicates the concentration of energy on $k$th frequency components where $k$ are multiples of $K$.
\end{IEEEproof}

\begin{corollary}
\label{cor:3}
The random permutation can eliminate the energy concentration on $k$th frequency components where $k\neq0$ are multiples of $K$.
\end{corollary}
\begin{IEEEproof}
If we randomly permute the orders of data sequences for exchangeable data blocks, we see (\ref{eq_pro3}) no longer holds except for the first case $n<B$. As a result, $F_k$ cannot be factorized into the product of two summations as previously shown in (\ref{eq_pro4}), which implies that energy cannot be concentrated on a few frequency components.

However, when $k=0$ (i.e., DC component), $F_0=\sum_{n=0}^{KB-1}x_n=K\widetilde{F}_0$, which coincides with the result in (\ref{eq_pro6}). This is because a permutation in the order of data sequence does not affect the result of summation thanks to the commutativity of addition.
\end{IEEEproof}

Proposition~\ref{pro:3} indicates that skipping the random permutation may result in spikiness in some frequency components due to the energy concentration; whereas Corollary~\ref{cor:3} can smooth out this spikiness by spreading energy all over frequency components. In fact, for a typical spectral analysis where the number of values in the analysis is large enough, $K$ tends to a large number and therefore artifacts in the frequency domain only appear at higher frequencies.

\section{Experimental Results} \label{sec:experimental}
In lossy compression, coding schemes can be evaluated with two common criteria: compression ratio and reconstruction quality. Since the reconstruction quality of IDEALEM cannot be directly assessed using a conventional measure such as MSE and SNR, as discussed in Section~\ref{sec:lossy}, we need to utilize other measures. In particular, compelling reconstruction results were visually represented in our previous work~\cite{lee2016novel,wu2017statistical}. To corroborate the ability of IDEALEM to preserve reconstruction quality, we employ various measures for describing time series data available in data mining literature~\cite{mori2016similarity}. Here it is important not to lose \emph{significant patterns} in the original data, which could be abnormal or singular, requiring attention of data analysts.

We use a data set of power grid monitoring data from $\upmu$PMUs installed on-site at Lawrence Berkeley National Laboratory (LBNL). The data set is from two different $\upmu$PMUs (A6BUS1 and BANK514), which contains about half a month records of power grid monitoring data. Each $\upmu$PMU monitors three-phase measurements (1, 2, 3) of voltages (L) and currents (C), where the measurements, also known as \textbf{phasors}, are composed of the \textbf{magnitude} (MAG) and the \textbf{phase angle} (ANG) of sine waves in electricity~\cite{emmaUPMU,wen2015phase}.\footnote{For instance, `A6BUS1C1MAG' denotes the magnitude of current on phase 1 measured by a $\upmu$PMU named A6BUS1.} Among this data set, we only show experimental results on phase 1 data, as results are similar across different phases. Each time series occupies nearly 1~GB in binary representation.

\subsection{Compression Ratio Comparison} \label{sec:comparison}
We present the compression ratio of IDEALEM with the ratios of other floating-point compression methods presented in Section~\ref{sec:floating}. Table~\ref{table_one} shows the compression ratios of four compression methods for $\upmu$PMU data. ZFP used a tolerance parameter in the fixed-accuracy mode (option \texttt{-a 8})~\cite{LindstromZFP}, where the tolerance parameter specifies the maximum absolute difference between an uncompressed value and a reconstructed value.

\begin{table}[!t]
\renewcommand{\arraystretch}{1.0}
\caption{Compression Ratios}
\label{table_one}
\centering
\begin{tabular}{ccccc}
\hline
\textbf{Data/Compression} & ZFP & ISABELA & SZ & IDEALEM \\
\hline
A6BUS1C1MAG & 9.99 & 5.57 & 16.36 & 242.3 \\
A6BUS1L1MAG & 6.40 & 5.57 & 58.15 & 120.0 \\
BANK514C1MAG & 7.50 & 5.56 & 14.38 & 248.4 \\
BANK514L1MAG & 8.00 & 5.57 & 41.71 & 156.5 \\
A6BUS1C1ANG & 8.76 & 5.36 & 67.25 & 86.89 \\
A6BUS1L1ANG & 8.78 & 5.38 & 177.4 & 84.32 \\
BANK514C1ANG & 8.76 & 5.36 & 59.13 & 96.39 \\
BANK514L1ANG & 8.78 & 5.38 & 210.8 & 85.05 \\
\hline
\end{tabular}
\end{table}

ISABELA used the window size 512, the number of coefficients 15, the error rate 5, and the BSplines switch~\cite{ISABELASW}. Here, the error rate is a relative error bound (\%) per data point. SZ used the relative error bound and the relative bound ratio was set to 0.001~\cite{SZSW}, where the error bound and its ratio works on the global data value range, not per data point.

Table~\ref{table_one} also shows the compression ratio of IDEALEM, where we used $D=255$ and $\alpha=0.01$. For MAG data, we employed the standard mode (Section~\ref{sec:encoding_standard}) and set $B=32$; for ANG data, we employed the residual mode (Section~\ref{sec:encoding_residual}) and set $B=112$. We use different block sizes $B$ for MAG and ANG data due to different theoretical upper bounds of achievable compression ratios, as explained in Section~\ref{sec:fundamental}.

In Table~\ref{table_one}, we can see that for many $\upmu$PMU data, compression ratios reach close to the maximum achievable compression ratios: 256 for the standard mode and 99.56 for the residual mode. The results of delta mode with the same parameters are 99.19 for A6BUS1C1ANG, 38.21 for A6BUS1L1ANG, 99.21 for BANK514C1ANG, and 62.99 for BANK514L1ANG~\cite{lee2017improving,lee2017expanding}. Apparently, current phase angle data are more amenable to the delta transformation method, whereas voltage phase angle data to the residual transformation method. This shows that depending on data types, one method can show better compression performance and there is no clear winner in transformation methods.

\subsection{Reconstruction Quality Comparison} \label{sec:reconstruction}
We present the reconstruction quality with the following measures: (\#1) number of local maxima (peaks), (\#2) mean distance between peaks, (\#3) mean distance between the values of peaks, (\#4) mean size of jumps, (\#5) number of jumps higher than 10\% of difference between the maximum and minimum values of series, and (\#6) the percentage of points that lie outside the whiskers of the Tukey box plot (more than 1.5 interquartile ranges above the upper quartile or below the lower quartile). With the same parameter configurations used in Table~\ref{table_one}, Table~\ref{table_two} shows the reconstruction qualities of all compression methods along with original data in terms of these six measures.

\begin{table*}[!t]
\renewcommand{\arraystretch}{1.0}
\caption{Reconstruction Qualities with Six Measures}
\label{table_two}
\centering
\setlength\tabcolsep{2.5pt}
\begin{tabular}{c|ccccc|ccccc}
\hline
\textbf{Data/Compression} & \#1 & ZFP & ISABELA & SZ & IDEALEM & \#2 & ZFP & ISABELA & SZ & IDEALEM \\
\hline
A6BUS1C1MAG & 20618980 & 1493207 & 20678257 & 18423314 & 41187683 & 6.04 & 83.34 & 6.02 & 6.75 & 3.02 \\
A6BUS1L1MAG & 20325437 & 1834458 & 20393068 & 7768430 & 41059933 & 6.12 & 67.84 & 6.10 & 16.02 & 3.03 \\
BANK514C1MAG & 16198077 & 10951716 & 16829379 & 13386853 & 41154067 & 7.68 & 11.36 & 7.39 & 9.30 & 3.02 \\
BANK514L1MAG & 17308827 & 251624 & 17410171 & 9991781 & 41108185 & 7.19 & 488.7 & 7.15 & 12.45 & 3.03 \\
A6BUS1C1ANG & 21591680 & 857406 & 22104731 & 7493261 & 22281503 & 5.76 & 145.1 & 5.63 & 16.61 & 5.59 \\
A6BUS1L1ANG & 622017 & 41255 & 1509371 & 34495 & 1112170 & 200.1 & 3016.3 & 82.44 & 3607.5 & 111.9 \\
BANK514C1ANG & 16148705 & 1934105 & 16811692 & 7620889 & 17017975 & 7.71 & 64.33 & 7.40 & 16.33 & 7.31 \\
BANK514L1ANG & 1561848 & 48359 & 2430822 & 89356 & 1456997 & 79.66 & 2572.9 & 51.19 & 1392.4 & 85.40 \\
\hline
\textbf{Data/Compression} & \#3 & ZFP & ISABELA & SZ & IDEALEM & \#4 & ZFP & ISABELA & SZ & IDEALEM \\
\hline
A6BUS1C1MAG & 0.27 & 0.23 & 0.27 & 0.28 & 0.23 & 0.35 & 2.14 & 0.36 & 0.40 & 0.48 \\
A6BUS1L1MAG & 0.46 & 0.71 & 0.45 & 0.61 & 0.45 & 0.44 & 2.88 & 0.46 & 1.15 & 0.96 \\
BANK514C1MAG & 5.64 & 7.79 & 5.54 & 6.26 & 5.85 & 9.86 & 14.98 & 9.66 & 11.93 & 11.54 \\
BANK514L1MAG & 0.03 & 0.00 & 0.03 & 0.03 & 0.03 & 0.04 & 2.00 & 0.04 & 0.07 & 0.06 \\
A6BUS1C1ANG & 0.91 & 19.41 & 1.02 & 1.80 & 0.91 & 1.25 & 26.55 & 1.51 & 3.60 & 0.87 \\
A6BUS1L1ANG & 3.80 & 123.3 & 7.67 & 47.82 & 7.63 & 9.58 & 145.2 & 5.75 & 172.7 & 6.46 \\
BANK514C1ANG & 1.53 & 10.81 & 1.65 & 2.17 & 1.23 & 1.68 & 12.31 & 1.99 & 3.55 & 0.97 \\
BANK514L1ANG & 2.13 & 106.5 & 4.84 & 23.95 & 5.72 & 3.85 & 124.6 & 3.62 & 67.31 & 4.37 \\
\hline
\textbf{Data/Compression} & \#5 & ZFP & ISABELA & SZ & IDEALEM & \#6 & ZFP & ISABELA & SZ & IDEALEM \\
\hline
A6BUS1C1MAG & 67644 & 95868 & 60379 & 66764 & 833079 & 0.00 & 0.09 & 0.00 & 0.00 & 0.01 \\
A6BUS1L1MAG & 32 & 37 & 28 & 33 & 686 & 0.32 & 0.21 & 0.32 & 0.33 & 0.32 \\
BANK514C1MAG & 1062126 & 1070228 & 1079142 & 1058476 & 2415159 & 3.16 & 3.16 & 3.16 & 3.16 & 2.80 \\
BANK514L1MAG & 39 & 251621 & 37 & 42 & 9824 & 0.11 & 0.00 & 0.11 & 0.12 & 0.12 \\
A6BUS1C1ANG & 50755 & 67443 & 50755 & 50834 & 27152 & 0 & 0 & 0 & 0 & 0 \\
A6BUS1L1ANG & 19880 & 19271 & 20618 & 19447 & 25015 & 0 & 0 & 0 & 0 & 0 \\
BANK514C1ANG & 45882 & 57293 & 45882 & 45980 & 23277 & 0 & 0 & 0 & 0 & 0 \\
BANK514L1ANG & 19904 & 19561 & 20569 & 19752 & 22959 & 0 & 0 & 0 & 0 & 0 \\
\hline
\end{tabular}
\end{table*}

Note that in Table~\ref{table_two}, smaller difference between the first column of each measure and results of compression methods indicates better reconstruction quality. For (\#1), ISABELA performs the best for the most part. The reconstruction quality of IDEALEM for MAG data is unsatisfactory due to the random permutation in the standard mode. While the random permutation avoids artificial patterns and eliminates the energy concentration on frequency components, it also yields high frequency components that contribute to numerous local peaks. However, IDEALEM shows good reconstruction quality for ANG data comparable to that of ISABELA. For (\#2), (\#3), and (\#4), IDEALEM is the second best compression method after ISABELA. Specifically, the large numbers of local peaks for reconstructed MAG data observed in (\#1) lead to small distances between them for the reconstructed MAG data of IDEALEM in (\#2). While ISABELA shows very good reconstruction qualities here, it is largely attributable to the non-aggressive compression characteristics of ISABELA as shown in Table~\ref{table_one}.

For (\#5), SZ shows overall the best performance. On the other hand, IDEALEM performs poorly in most cases. This is again because of numerous local peaks observed in (\#1), which directly corresponds to a large number of jumps. However, for (\#6), IDEALEM is comparable with ISABELA and SZ. In particular, (\#6) measures the percentages of outliers in data set, which indicates whether reconstructed data can preserve the original value range of the data set. Since ANG data has the bounded range of $0^{\circ}$ to $360^{\circ}$, outliers do not exist in these results. All in all, we note that IDEALEM preserves reconstruction quality with the exception of measures affected by the number of peaks/jumps. Considering its very high compression ratios, the results shown here confirms the effectiveness of IDEALEM as a new lossy compression method.

\subsection{Spectral Analysis} \label{sec:spectral}
\begin{figure*}[!t]
\centering
\subfloat[A6BUS1C1MAG Original]{\includegraphics[width=0.5\linewidth]{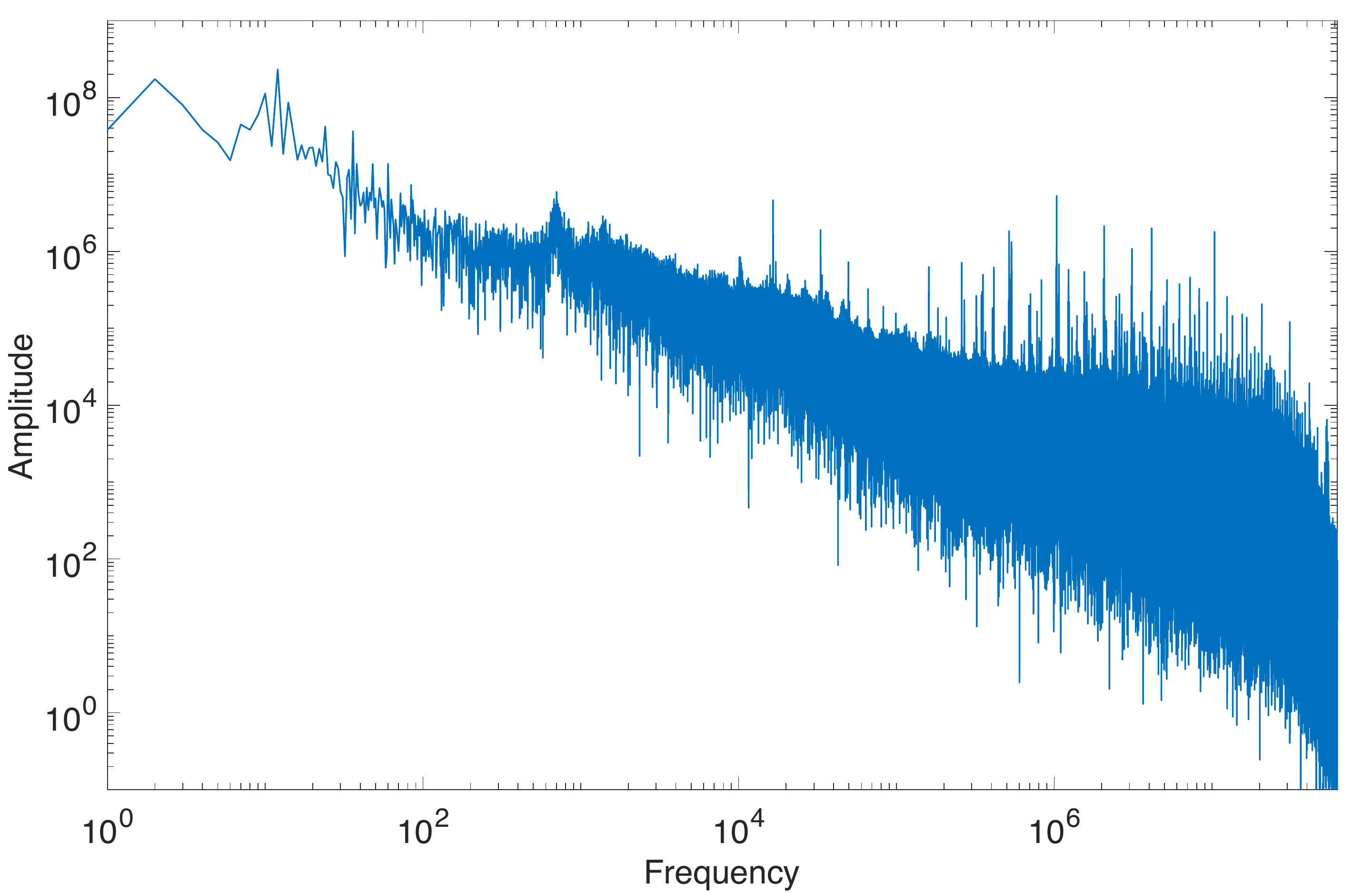}%
\label{fig_twelve_a}}
\hfil
\subfloat[A6BUS1C1MAG Reconstructed]{\includegraphics[width=0.5\linewidth]{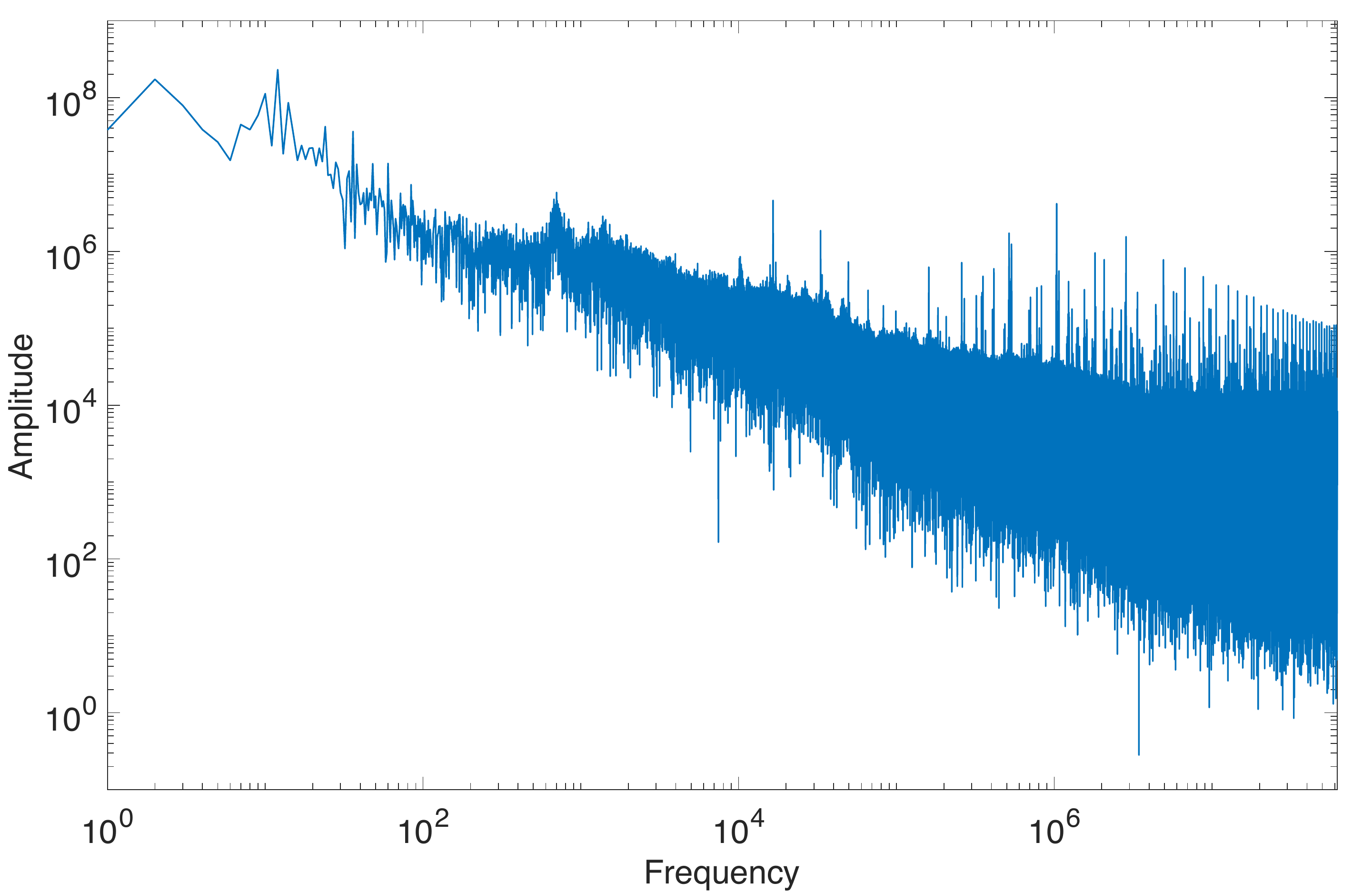}%
\label{fig_twelve_b}}
\\
\subfloat[A6BUS1C1ANG Original]{\includegraphics[width=0.5\linewidth]{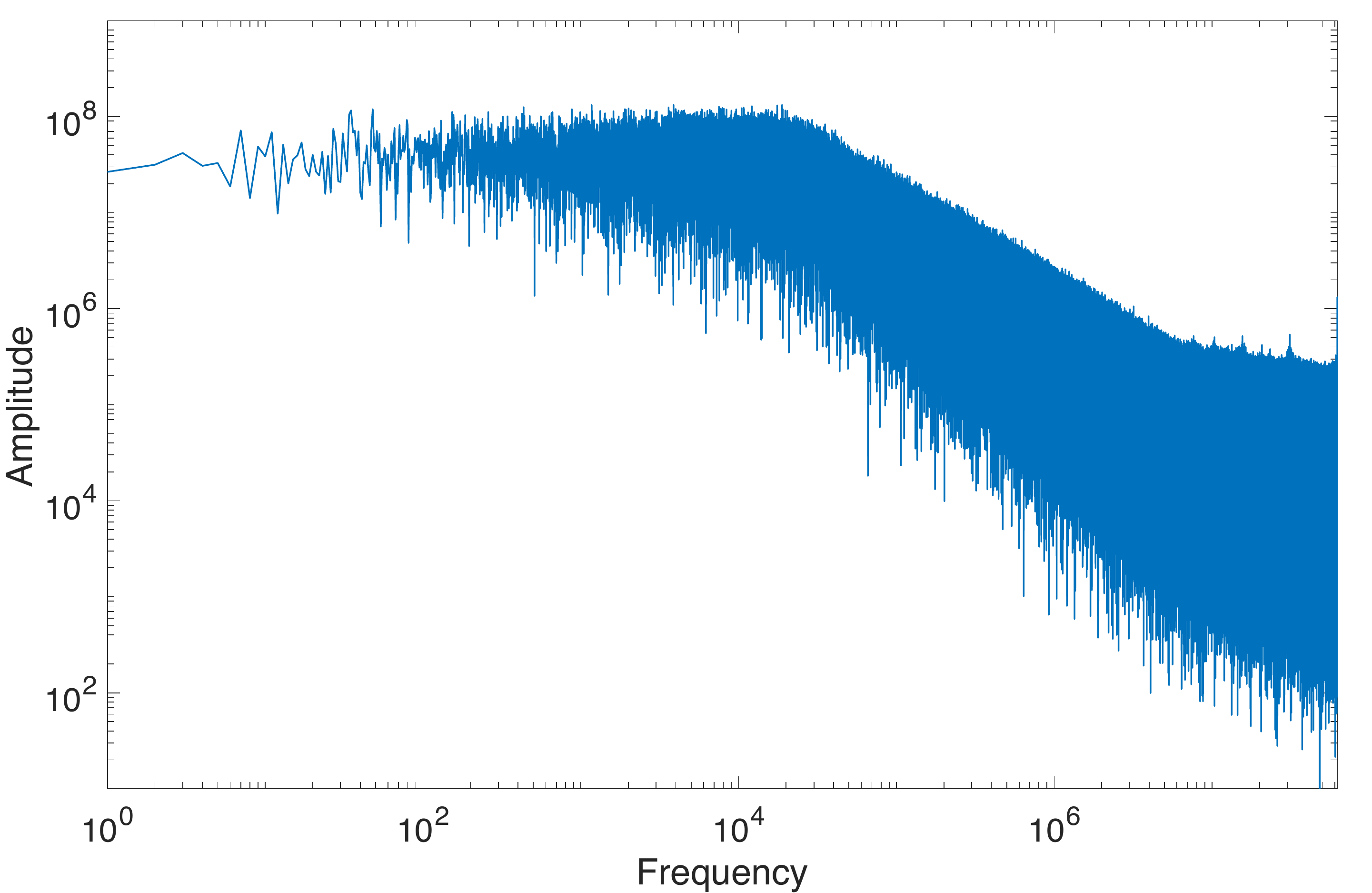}%
\label{fig_twelve_c}}
\hfil
\subfloat[A6BUS1C1ANG Reconstructed]{\includegraphics[width=0.5\linewidth]{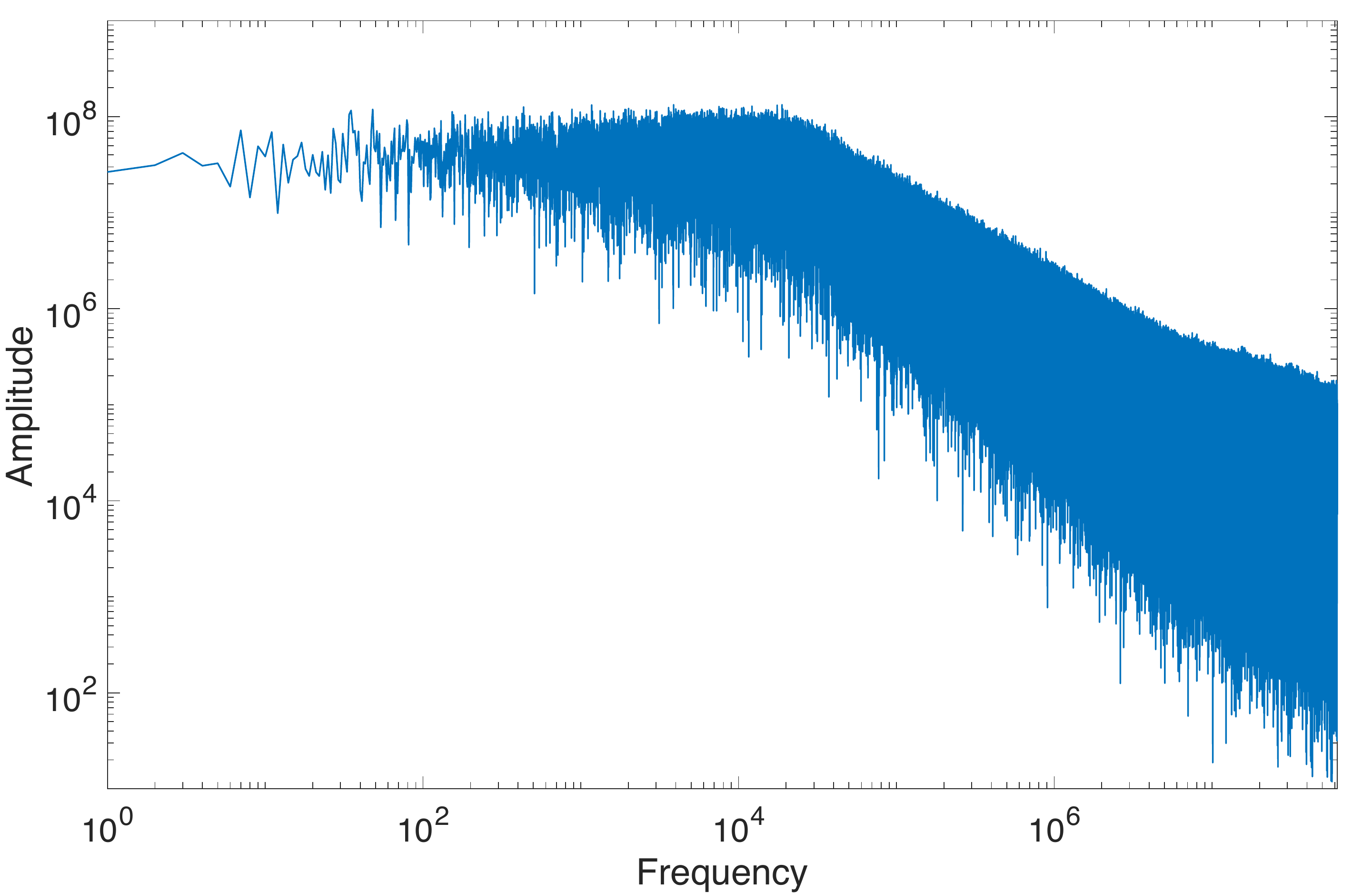}%
\label{fig_twelve_d}}
\caption{Amplitude spectra of DFT for A6BUS1C1MAG and A6BUS1C1ANG, where both axes are drawn in log scale. Original data and its IDEALEM reconstruction are shown together. The Fourier spectrum of IDEALEM reconstruction is close to that of the original especially at lower frequencies.}
\label{fig_twelve}
\end{figure*}

For the spectral analysis of IDEALEM, we perform the DFT of original and reconstructed data and compare their amplitude spectra in Fig.~\ref{fig_twelve}. In particular, the spectra of A6BUS1C1MAG and A6BUS1C1ANG are shown here to represent MAG and ANG data respectively. Note that we only show the single-sided spectrum here since the amplitude of DFT is completely symmetric. In addition, the DC component is excluded from each plot, as its amplitude is so high that it can hinder the analysis of other components.

In Fig.~\ref{fig_twelve}, we can see that the reconstruction quality of IDEALEM is good at lower frequencies, which implies important frequency components are well preserved. For many application domains including power grid monitoring, low frequency components are considered more important than high frequency components since high frequency components are many orders of magnitude smaller than low frequency components.

As for high frequency components, the IDEALEM reconstruction tends to increase their amplitudes in the standard mode due to the random permutation as shown in Fig.~\ref{fig_twelve_b}, which explains the results of (\#1) and (\#2) in Table~\ref{table_two}. On the other hand, Fig.~\ref{fig_twelve_d} does not show any amplitude increases in those components, as the residual mode does not employ the random permutation.

\begin{figure}
\centering
\includegraphics[width=0.7\columnwidth]{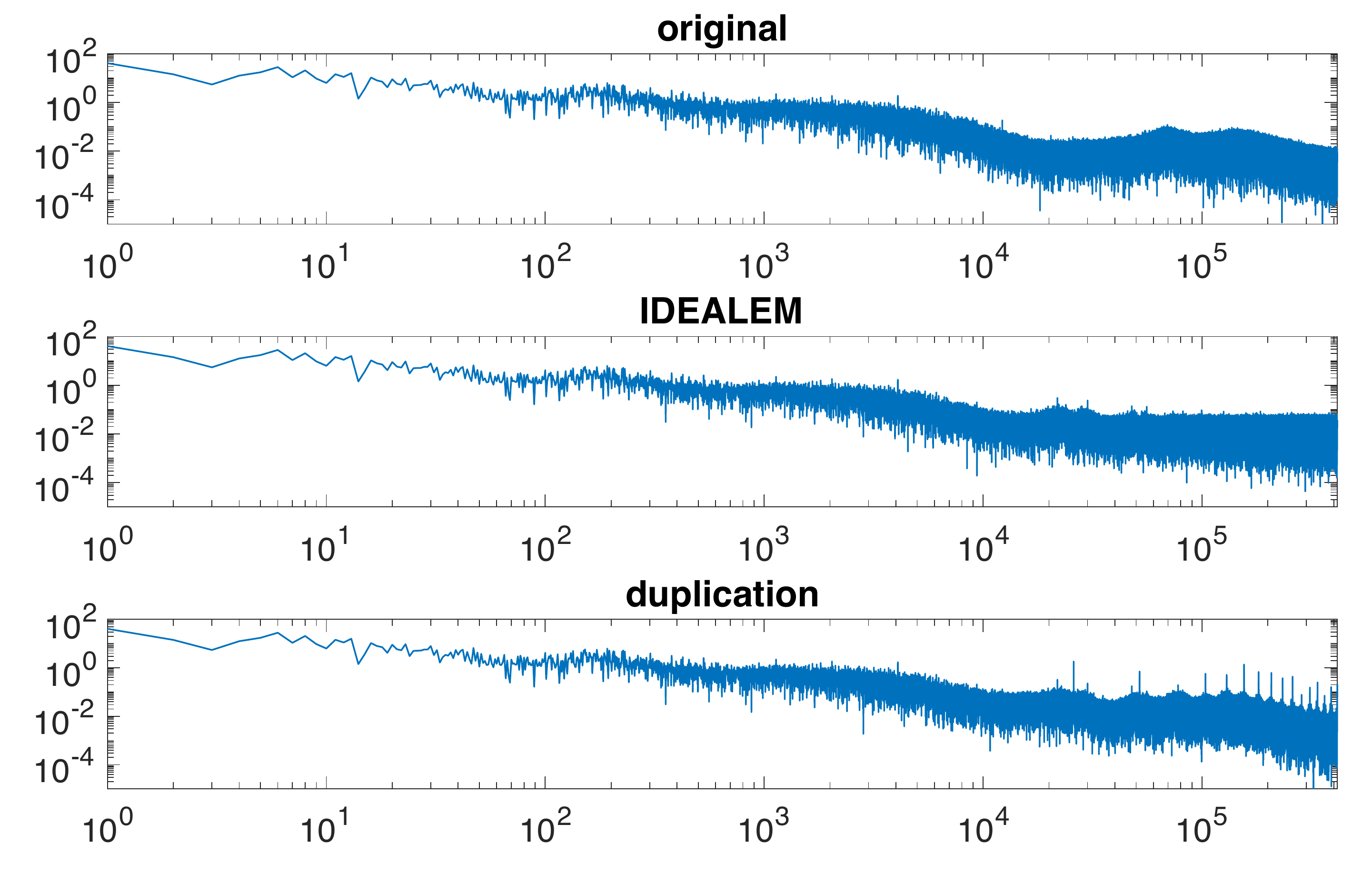}
\caption{Amplitude spectra of DFT for the intracranial EEG of rodent, where both axes are drawn in log scale. Original data, IDEALEM reconstruction, and no random permutation (duplication) are shown together. The duplication creates energy concentrations shown as spikes at higher frequencies.}
\label{fig_thirteen}
\end{figure}

In order to closely observe how the random permutation affects the Fourier spectrum, we use another data set of intracranial electroencephalography (EEG)~\cite{ledochowitsch2015strategies,wu2017statistical}, which consists of 830,000 values from a single channel EEG recording from a rodent. Fig.~\ref{fig_thirteen} shows three single-sided amplitude spectra of DFT for original data, IDEALEM reconstruction, and no random permutation during the reconstruction (duplication).

We can observe that in Fig.~\ref{fig_thirteen}, the duplication shows spikes at higher frequencies while the amplitudes of high frequency components in general, apart from these spikes, are not boosted. On the other hand, the IDEALEM reconstruction does not show any spikes while its high frequency components are amplified. These two observations confirm Proposition~\ref{pro:3} and Corollary~\ref{cor:3}, respectively.

\subsection{Effectiveness of Min/Max Check}
\begin{figure*}[!t]
\centering
\subfloat[\#1 MAG Data]{\includegraphics[width=0.5\linewidth]{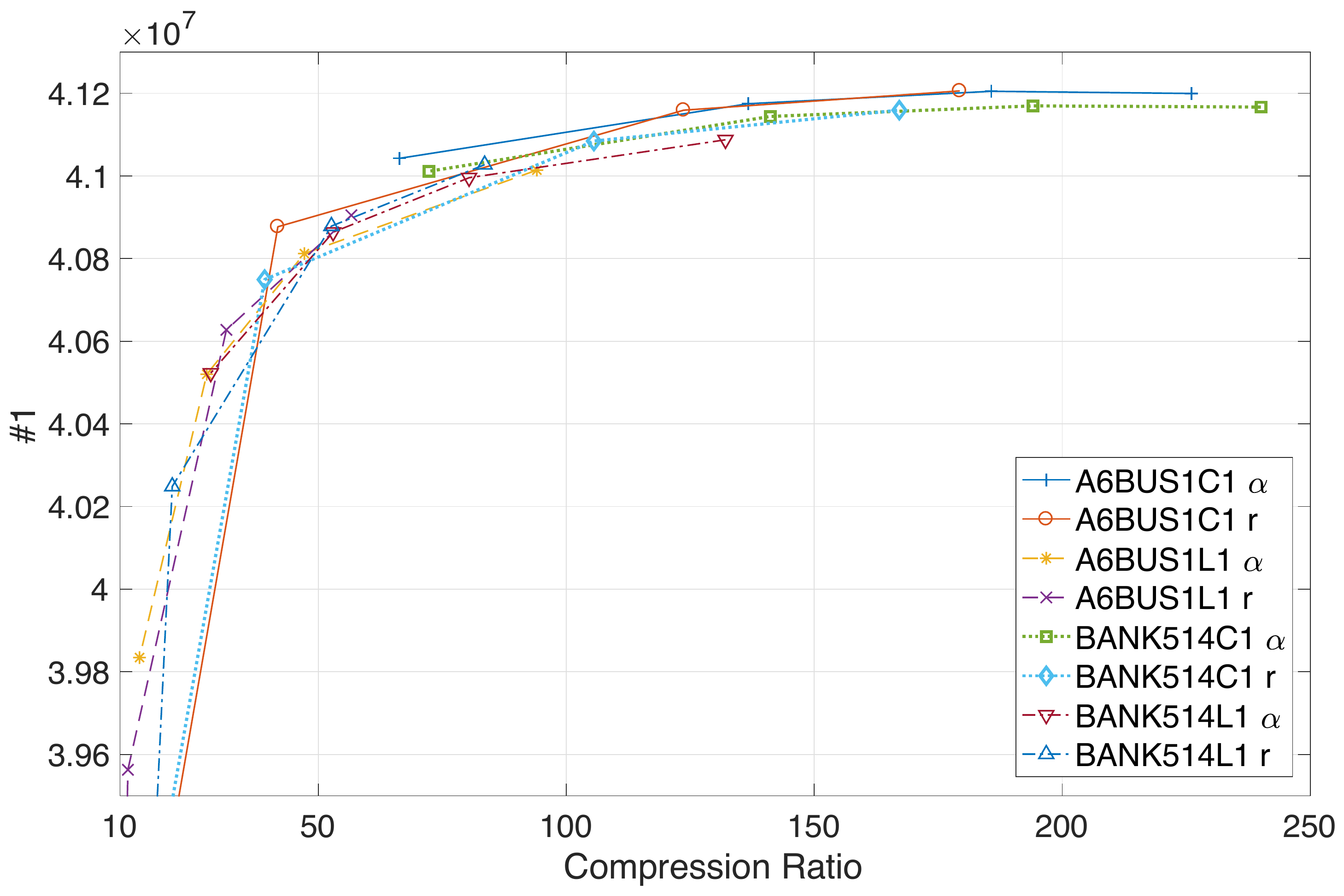}%
\label{fig_fourteen_a}}
\hfil
\subfloat[\#1 ANG Data]{\includegraphics[width=0.5\linewidth]{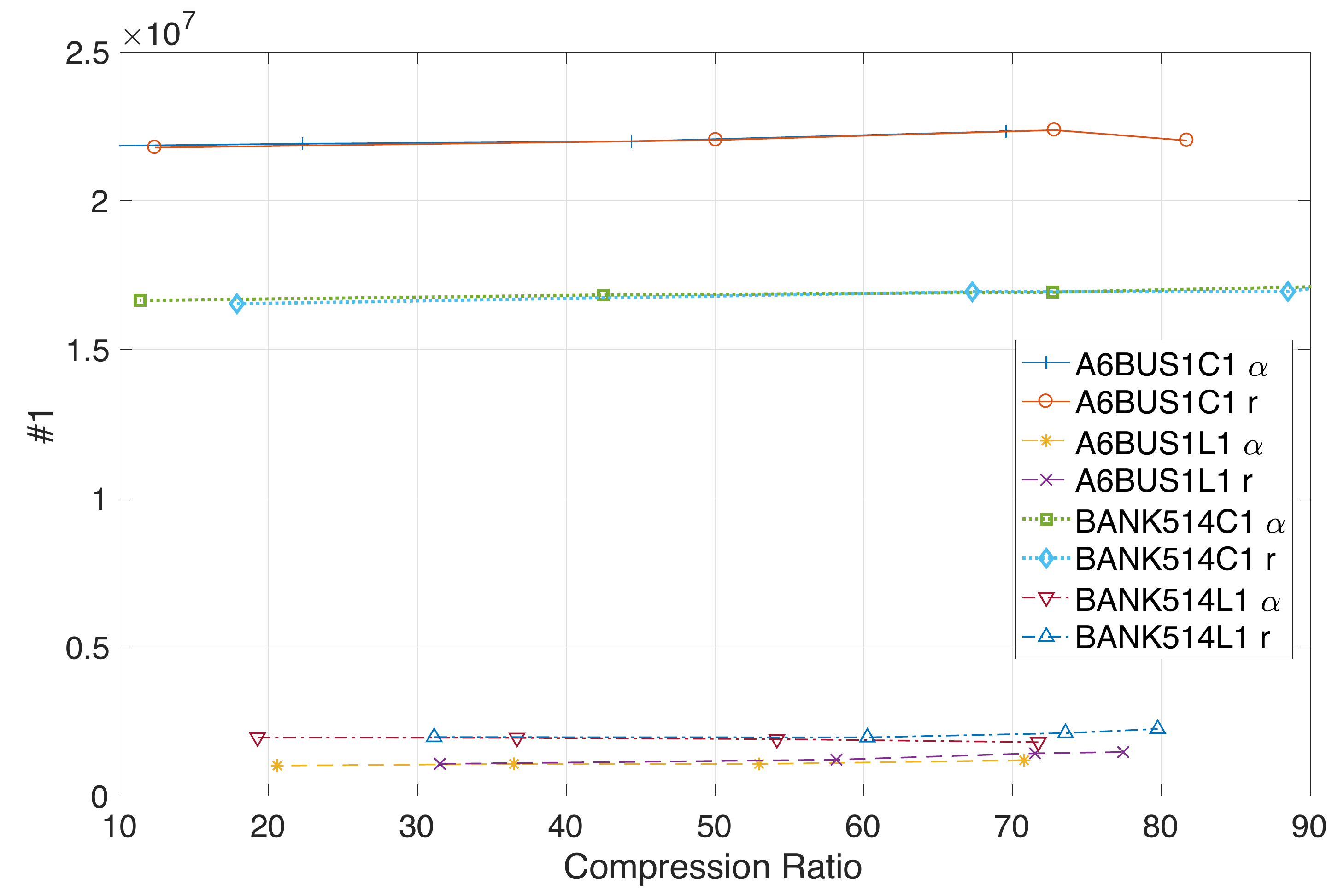}%
\label{fig_fourteen_b}}
\\
\subfloat[\#5 MAG Data]{\includegraphics[width=0.5\linewidth]{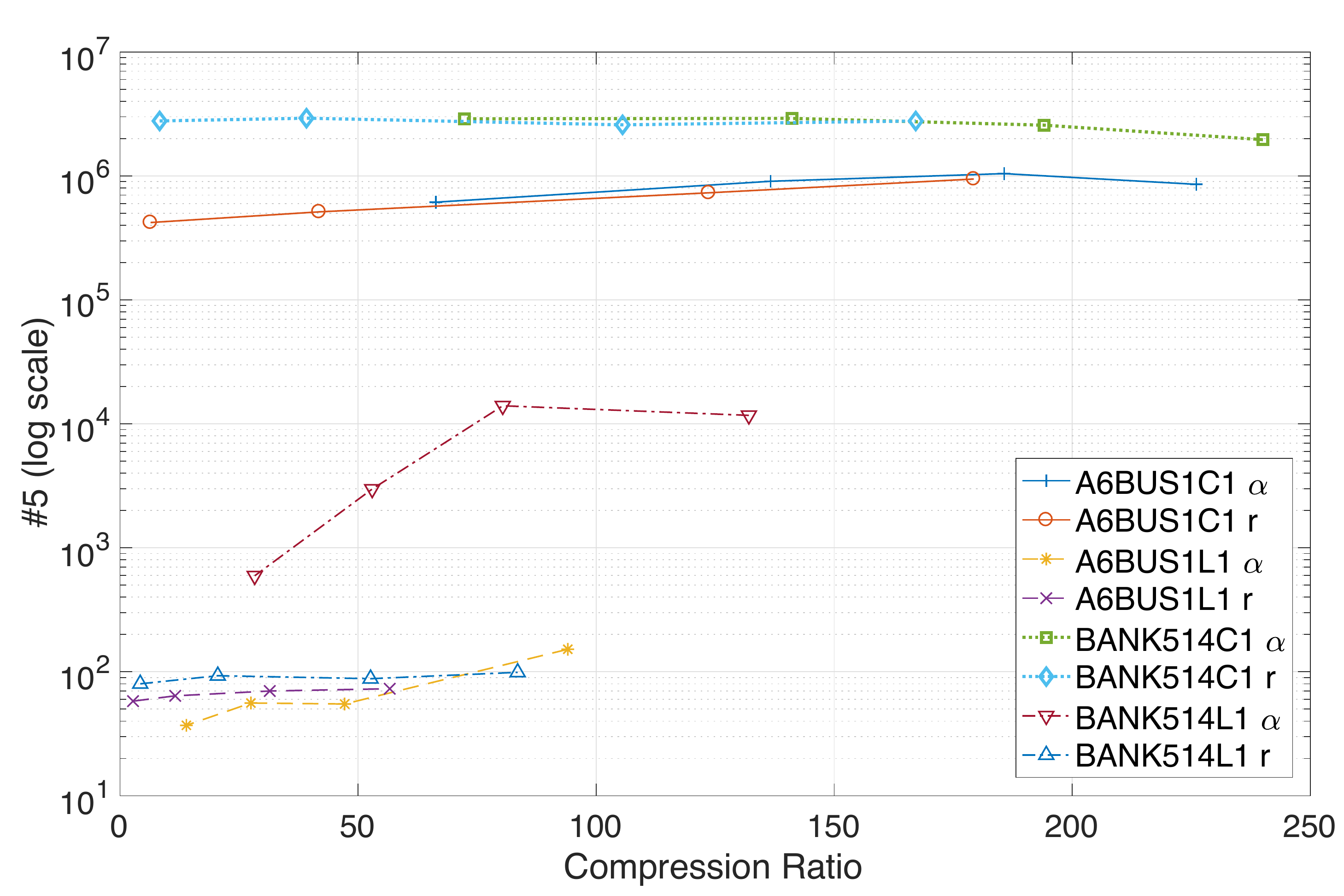}%
\label{fig_fourteen_c}}
\hfil
\subfloat[\#5 ANG Data]{\includegraphics[width=0.5\linewidth]{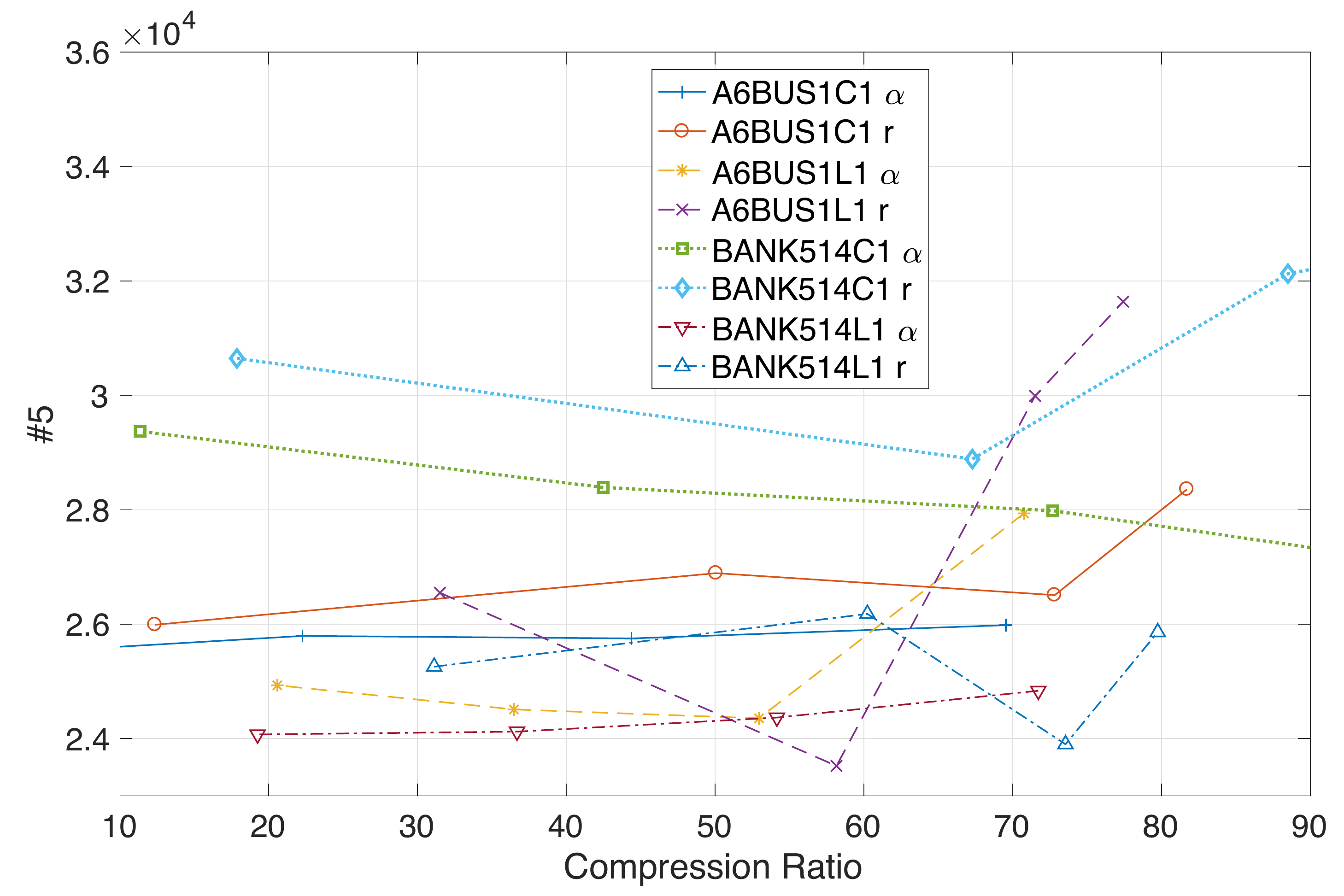}%
\label{fig_fourteen_d}}
\caption{Reconstruction quality versus compression ratio for parameter changes with $\alpha$ and $r$. Both parameters yield similar results in most cases, except for (c) where using $r$ is more advantageous to BANK514L1MAG. Note that the y-axis is drawn in log scale for (c).}
\label{fig_fourteen}
\end{figure*}

In Section~\ref{sec:minmax}, we discussed the min/max check that can complement the KS~test. IDEALEM controls the compression ratio and reconstruction quality by the threshold $\alpha$ (KS~test) and/or the relative tolerance $r$ (min/max check). Fig.~\ref{fig_fourteen} shows the reconstruction quality plotted against the compression ratio with these two parameters. In particular, quality measures (\#1) and (\#5) in Table~\ref{table_two} are presented with parameter changes $\alpha=0.02,0.05,0.1,0.2$ and $r=0.1,0.2,0.3,0.4$ for a fixed $\alpha=0.01$.

In Fig.~\ref{fig_fourteen}, we observe that IDEALEM in general yields similar pairs of compression ratio and reconstruction quality. However, for the quality measure (\#5) shown in Fig.~\ref{fig_fourteen_c}, the result of BANK514L1MAG is improved much when we tune IDEALEM with $r$. In Table~\ref{table_two}, the quality measure (\#5) of original BANK514L1MAG is 39, and adjusting $r$ instead of $\alpha$ produces results closer to 39 within a similar compression ratio range.

\begin{figure}
\centering
\includegraphics[width=0.6\columnwidth]{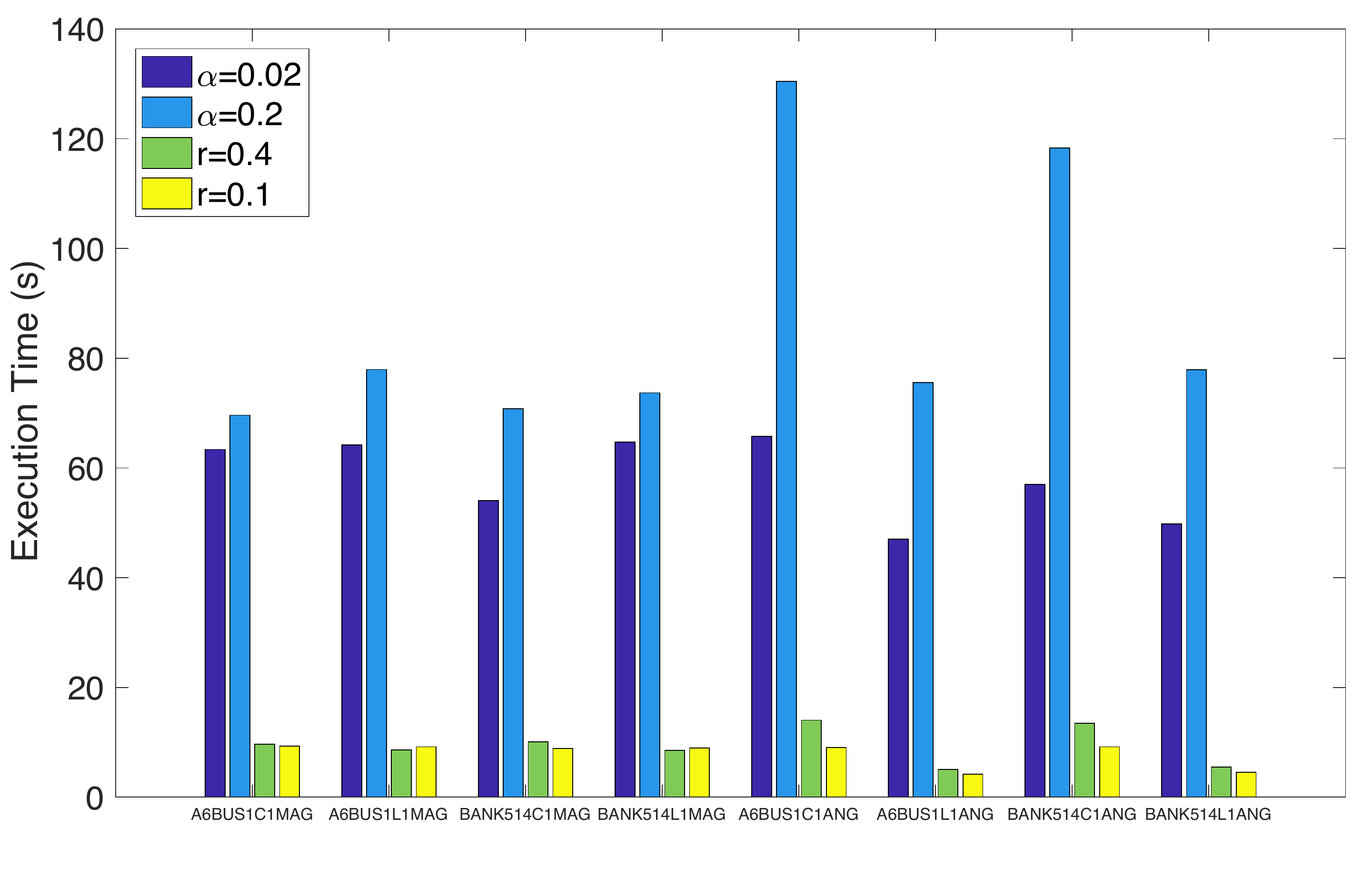}
\caption{Execution time of encoder measured in seconds for parameter changes with $\alpha$ and $r$. Using the min/max check is many times faster than using the KS~test alone. Note that $\alpha=0.2$ takes more time than $\alpha=0.02$ because it is more difficult to pass the KS~test with a higher $\alpha$, which leads to a longer time searching the buffer. On the other hand, encoding time is similar for different $r$.}
\label{fig_fifteen}
\end{figure}

Aside from this quality improvement, it is noteworthy that the min/max check can significantly accelerate the encoding process of IDEALEM. Fig.~\ref{fig_fifteen} shows the execution time of encoder for our data set averaged over 100 trials. Experiments were conducted on a laptop equipped with Intel Core i7 (2.7 GHz) CPU and 16 GB RAM, which runs macOS 10.12.6 (Darwin kernel 16.7.0). Note that the execution time of decoder is the same for both the KS~test only and the min/max check cases, and the decoding process of IDEALEM is faster than encoding since decoding does not need to search the buffer~\cite{lee2016novel}.

In Fig.~\ref{fig_fifteen}, we can observe a large difference in the execution time between adjusting $\alpha$ and adjusting $r$. Without the min/max check, the KS~test should be repetitively performed across dictionary blocks in the buffer until a similar distribution is found. However, having the min/max check prior to the LEM processing filters out lots of data blocks, which decreases the overall number of KS~tests performed in the encoding process.

It should be noted that in Fig.~\ref{fig_fifteen}, $\alpha=0.2$ takes longer execution time than $\alpha=0.02$. Increasing $\alpha$ is equivalent to raising the bar for similarity; thus fewer data blocks becomes exchangeable, which leads to a lower compression ratio. This prolongs buffer search time and therefore overall execution time.

On the other hand, the encoding time remains similar for $r=0.4$ and $r=0.1$, while decreasing $r$ here is equivalent to lowering the tolerance, which also leads to a lower compression ratio. In fact, in many cases, decreasing $r$ yields shorter execution time and this difference in time becomes more pronounced when there is a large time difference between different $\alpha$ (e.g., A6BUS1C1ANG and BANK514C1ANG). Unlike the LEM processing by KS~test, the min/max check is very fast and filtering out more data blocks does not impact the execution time, but help IDEALEM to reduce the burden of LEM processing in many cases.

Finally, we discuss the effectiveness of min/max check in terms of preserving significant patterns in the original data that require the attention of data analysts. In power grid monitoring, an analyst often monitors changes in voltage by a \textbf{tap changer}, which provides $\pm$10\% voltage adjustment in thirty-two 0.625\% steps~\cite{siemens2017}. For example, if the nominal voltage is 7,200~V, one may monitor tap changes that are greater than or equal to 45~V (0.625\%) between sample values.

In our experiments with four days records of data containing frequent tap changes, some of changes were difficult to detect with the KS~test only encoding (even with high $\alpha$), because these changes lasted very briefly. On the contrary, all the tap changes were successfully detected with the min/max check encoding: even a high tolerance $r=0.5$ with a low $\alpha=0.01$ could subsequently produce reconstructed data where all the tap changes remained intact.

\section{Conclusion} \label{sec:conclusion}
We have proposed IDEALEM that leverages key statistical properties of the original data and reduces data size needed to keep common and uninteresting data records. IDEALEM handles both stationary and non-stationary time series, where non-stationary time series data is handled by the residual/delta transformation. IDEALEM also employs the min/max check, which significantly reduces execution time for encoding, that can better capture important features lasting only for a brief duration.

We provided a comprehensive analysis on very high compression performance of IDEALEM in the standard mode (stationary data) and the residual/delta mode (non-stationary data). In order to study its reconstruction quality, we employed various quality measures for describing time series data and found that IDEALEM shows good reconstruction quality. Spectral analysis also showed the reconstruction quality is not compromised in terms of important frequency components for application domain.

Currently, IDEALEM is designed to handle one-dimensional array of data, which inherently ignores correlated data generated from different devices. We plan to expand IDEALEM into a multidimensional compression algorithm that leverages the joint correlation. To this end, we want to explore various multivariate statistical similarity measures to significantly improve the compression ratio and reconstruction quality of IDEALEM.

% or


%
%\appendices
%% you can choose not to have a title for an appendix
%% if you want by leaving the argument blank
%\section{}
%% use section* for acknowledgment
%\ifCLASSOPTIONcompsoc
%  % The Computer Society usually uses the plural form
%\else
%  % regular IEEE prefers the singular form
%\fi

\ifCLASSOPTIONcaptionsoff
  \newpage
\fi



\bibliographystyle{IEEEtran}
\bibliography{references}
\end{document}